\documentstyle[11pt]{article}

\makeatletter
\long\def\@makecaption#1#2{
 \vskip 10pt
 \setbox\@tempboxa\hbox{{\small\bf#1:} \small#2}
 \ifdim \wd\@tempboxa >\hsize {\small\bf#1:} \small#2\par
 \else \hbox to\hsize{\hfil\box\@tempboxa\hfil}
 \fi}
\makeatother

\def\bigfigsize{14 truecm}
\def\bigcapsize{12 truecm}
\def\medfigsize{9 truecm}
\def\medcapsize{4.8 truecm}

\def\smallfigsize{6.9 truecm}
\def\smallcapsize{6 truecm}

\def\twofigsize{6.9 truecm}
\def\twocapsize{6 truecm}

\font\smalltitlefont = cmss10 scaled \magstep 3
\font\sectionfont    = cmss10 scaled \magstep 2
\font\subsectionfont = cmss10 scaled \magstep 1

\makeatletter

\def\maketitle{
	\par
	\begingroup
		\def\thefootnote{\fnsymbol{footnote}}
		\def\@makefnmark{\hbox to 0pt{$^{\@thefnmark}$\hss}}
		\if@twocolumn \twocolumn[\@maketitle]
		\else
			\newpage
			\global\@topnum\z@ \@maketitle
		\fi
		\thispagestyle{plain}
		\@thanks
	\endgroup
	\setcounter{footnote}{0}
	\let\maketitle\relax
	\let\@maketitle\relax
	\gdef\@thanks{}
	\gdef\@author{}
	\gdef\@title{}
	\let\thanks\relax
	}

\def\@maketitle{
	\newpage
	\null
	\vskip 2em
	\begin{center}
		{\smalltitlefont \baselineskip 20pt \@title \par}
		\vskip 1.5em
		{
			\large \lineskip .5em
			\begin{tabular}[t]{c} \@author \end{tabular}
			\par
		}
		\vskip 1em
		{\large \@date}
	\end{center}
	\par
	\vskip 1.5em
	}

\def\section{\@startsection {section}{1}{\z@}{-3.5ex plus -1ex minus
 -.2ex}{2.3ex plus .2ex}{\sectionfont}}
\def\subsection{\@startsection{subsection}{2}{\z@}{-3.25ex plus-1ex
     minus-.2ex}{1.5ex plus.2ex}{\subsectionfont}}

\makeatother

\newcommand{\Zeta}{{\cal Z}}
\def\romanic#1{{\uppercase\expandafter{\romannumeral #1}}}

\makeatletter
\def\alt{\mathrel{\mathpalette\vereq<}}
\def\vereq#1#2{\lower3pt\vbox{\baselineskip1.5pt \lineskip1.5pt
	\ialign{$\m@th#1\hfill##\hfil$\crcr#2\crcr\sim\crcr}}}

\makeatother

\input epsf   

\def\today{\number\day \
     \ifcase\month\or Jan\or Feb\or Mar\or Apr\or May\or June
	 \or July\or Aug\or Sep\or Oct\or Nov\or Dec \fi \
     \number\year}

\begin{document}

\title{Thermal phenomenology of hadrons \\
   from 200 A GeV S+S collisions
   }

\author{Ekkard Schnedermann,$^{a,b}$ Josef Sollfrank$^a$ and Ulrich Heinz$^a$
   \\ [6 pt]
   $^{(a)}$Institut f{\"u}r Theoretische Physik, Universit{\"a}t
Regensburg\\[-2pt]
   D-93040 Regensburg, Germany \\
   and \\
   $^{(b)}$Physics Department, Brookhaven National Laboratory\\[-2pt]
   Upton, New York 11973, USA \\
   }
\date{\today}

\maketitle

\begin{abstract}
\noindent
We develop a complete and consistent description for the hadron spectra
from heavy ion collisions in terms of a few
collective variables, in particular temperature, longitudinal
and transverse flow.
To achieve a meaningful comparison with presently available data, we
also include the resonance decays into our picture. To disentangle the
influences of transverse flow and resonance decays in the $m_T$-spectra,
we analyse in detail the shape of the $m_T$-spectra.
\end{abstract}

\section{Introduction}
\label{introduction}

Our current understanding of QCD results basically from high energy
experiments with small collision systems suffering hard interactions
which are relatively
easy to analyse. In a first attempt to test QCD predictions for larger
systems, especially the predicted phase transition from hadronic matter
to the hypothetical quark gluon plasma, existing accelerators were modified
to experiment with nuclei instead of only protons. The first round of
experiments with nuclear beams took place during the years 1986--1990 at the
AGS (Alternating Gradient Synchrotron) of the Brookhaven National Laboratory
(BNL) and at the SPS (Super Proton Synchrotron) at CERN.
While the biggest possible projectile nuclei $^{28}Si$ (BNL) and
$^{32}S$ (CERN) do not deserve the title ``heavy-ions'', the situation
on the experimental side will considerably improve with the $Au$-beam
at the BNL, which is already in operation, and the planned $Pb$--beam
for CERN which is planned for 1994. To fully utilize the new possibilities
new methods for analysing the heavy-ion data have to be developed, which
are capable to characterize the physical situation over the whole range
of nuclei and for both the energies at BNL and CERN.

In this paper we want to develop a phenomenological model for the
hadronic matter by starting out with thermalization as the basic assumption
and adding more features as they are dictated by the analysis of the measured
hadronic spectra. Eventually we will show that under the assumption of
collective flow almost all hadronic spectra can be described by the model.
The model can be extended to a consistent hydrodynamical
description \cite{globalhydro}, which provides the link to the possible initial
conditions and can in turn be used to extract new information about the
final state of the hadronic system \cite{circumstantial,hydrodraft}.

We will start by explaining our choice of data in section
\ref{choiceofdata}. In section \ref{stationarythermalsource} we define
the stationary thermal model for particle emission and show its failure
for the measured $\pi^-$ $m_T$-spectra. We subsequently refine it by
introducing resonances and their decay contributions to the $m_T$-spectra
of all measured hadronic species which can then be described successfully by
a uniform temperature. The analysis of the rapidity
distributions in section \ref{longitudinalflow} leads us to the introduction
of longitudinal flow. On the other hand, as shown in section
\ref{transverseflow}, the existence of transverse flow
can not be established from the data.
However this can be clarified by a closer theoretical investigation
of the reaction prior to freeze-out, which was shortly reported in
\cite{circumstantial} and will be presented in detail in \cite{hydrodraft}.
Finally in section \ref{discussionandconclusions} we critically assess
the relevance of our model in the light of $pp$ data and give
a conclusion of our work.

\section{Choice Of Data}
\label{choiceofdata}

{}From the large amount of data from many experimental groups at the BNL
and at CERN, which have been measured with many different targets and
projectiles from $^{27}\!Al$ to $^{184}W$ \cite{otherdata}, we want to pick out
one set of data which is for our purpose easiest to interprete:
Transverse mass and rapidity distributions
of pions~\cite{NA35charged}, protons~\cite{NA35charged},
$K^0_s$, $\Lambda$ und $\overline \Lambda$~\cite{NA35strange}, as measured in
central collisions of $^{32}S$ with $S$ at $200\,\rm GeV/nucleon$
by the NA35 Collaboration at CERN with a streamer chamber.

For our selection we have the following reasons:
\begin{itemize}

\item
The high beam energy at CERN separates kinematically the
central reaction zone around rapidity $y_{\rm cm}=3$ from the fragmentation
regions of the target and projectile at $y=0$ and $y=6$. At BNL the
span between target and projectile is smaller ($y_{proj}-y_{lab}=3.4$),
and the width of each of the fragmentation regions $\Delta y \approx 1 - 2$
leads to a mixing of all zones.

\item
The higher energies at CERN also generate a stronger longitudinal expansion
of the matter, requiring at least cylindrical symmetry for the model,
which we will introduce in the next section, to describe the data. The BNL data
for the pions~\cite{Gersdorffdndy} are closer to an isotropic momentum
distribution and might be described well by a spherically symmetric model
as we have presented one already some time ago~\cite{Search}.

\item
The streamer chamber from NA35 provides for a large number of measurements
in the same experiment and thus guarantees easy comparibility. In this paper
we will analyse the spectra of pions, protons, $K^0_s$, $\Lambda$ and
$\overline \Lambda$, which form the bulk of the hadronic matter.

\item
The symmetric collision system $S$+$S$ together with the
selection of the events with $2\%$ of the highest multiplicities
minimizes the number of spectator nucleons,
which did not collide with other nucleons at all.
We are not interested in these nucleons,
because they are similar to the majority of the nucleons in $pA$ collisions,
which suffer only a small energy or momentum transfer each.
We are mainly interested in the participant nucleons, which
generate the highest energy densities and where we can expect
the strongest collective effects and would look for a quark gluon plasma.

\item
The kinematic symmetry is experimentally fortunate,
because only the wider open, easier accessible rear half of phase space
($y \le 3$) has to be measured, the other half can be obtained
by reflection around $y=3$. E.g.\ by measuring the negative hadrons
in the interval $0.8<y<3.0$
and reflecting around $y=3$, NA35 covers almost the whole rapidity gap between
projectile and target.
Also the isospin symmetry of $^{32}S$ can be employed for the equality of
proton and neutron spectra, as well as $\pi^-$, $\pi^0$ and $\pi^+$,
allowing for a determination of the proton spectra by subtracting the spectra
of negative tracks from the spectra of the positive tracks
with only a small error introduced
by the possible inequality of the $K^+$ and $K^-$ spectra and $\bar p$
contributions \cite{NA35charged}.

\end{itemize}

These advantages outweigh the disadvantage that $S$+$S$
is not the biggest possible collision system to date.
It contains with full overlap
at most 64 nucleons, whereas a bigger target, e.g.\ $^{197}\!Au$, would
increase the number of participating nucleons to 115 in a simple geometric
picture and moreover would raise the baryon and energy densities as well.
However, besides the blurring of the theoretical explanation
by the big number of non-participating nucleons,
for an asymmetric collision system the phase
space has also to be measured in forward direction, which is experimentally
increasingly difficult for the heavier collision systems. These data are thus
not available yet, $^{32}S$+$^{(107/109)}\!\!Ag$ is in the analysis right now
\cite{NA35TPC}. The biggest asymmetric system with a complete set of
published hadron data
is up to now $^{16}O+^{197}\!\!Au$ with a maximum number of participants
(approx. $70$ from geometry) again similar to $S$+$S$.

Since the fundamental dynamics of an ultrarelativistic heavy-ion reaction
will basically be determined by strong interactions, the regular hadrons
(pions, protons and kaons) carry the major part of the total
energy~\cite{Stroebele}. These particles spectra are most important for
the global picture of the reaction. To a large extent we will rely on the
pion spectra, because the pions are by number and also by energy
(2/3 of total)
the most important group and can be measured best, i.e.\ NA35 determines the
$\pi^-$ spectra by recording all negative tracks and only correcting them for
electron tracks. The contamination by other particle species, i.e. by
antiprotons or $K^-$, is quantitatively small and can be accounted for
on a statistical basis.

In this paper we will only be interested in the {\it shape} of the spectra.
The absolute multiplicities or the particle ratios
convey little direct information about the reaction dynamics
and require additional knowledge about the volume of the interaction zone
or the equation of state, respectively. We will defer these questions
to the more detailed hydrodynamical simulation of the collision zone
in \cite{hydrodraft}.

\section{Spectra From A Stationary Thermal Source}
\label{stationarythermalsource}

\subsection{A Purely Thermal Source}
\label{thermalsource}

In order to clarify the notation we start with the
invariant momentum spectrum of particles
radiated by a thermal source with temperature $T$:
\begin{equation}
E{{d^3n}\over{d^3p}} = {{dn}\over{dy\,m_T dm_T\,d\phi}}
= {{gV}\over{(2\pi)^3}} E \, e^{-(E-\mu)/T} \; ,
\end{equation}
where $g$ is the spin/isospin-degeneracy factor for the particle species
and $\mu$ the grand canonical potential $\mu=b\mu_b+s\mu_s$
as originating from its baryon and strangeness quantum numbers $b$ and $s$.
For simplicity we neglect quantum statistics with the reasoning that
its influence will be rather small at the low densities
where the particles typically decouple from each other and where
the spectra are computed. This also holds true for the pions,
whose Bose-character is only then important for the spectral shape
when chemical potentials close to the pion mass are
introduced \cite{pionpotential} via non-equilibrium arguments,
which are somewhat beyond the scope of our analysis.
$V$ is the volume of the source, giving together
with the factor $e^{\mu/T}$ the normalization of the spectrum,
which we will always adjust for a best fit to the data,
because we are only interested in the {\it shape} of the spectra to reveal
the dynamics of the collision zone at freeze-out.
In the remainder of the text we will always give the spectra
in terms of rapidity $y=\tanh^{-1} (p_L/E)$
and transverse mass  $m_T=\sqrt{m^2+p_T^2}$. Also $\hbar=c=k_B=1$.

We obtain the transverse mass spectrum $dn/(m_T dm_T)$ by integrating
over rapidity using the modified Bessel function $K_1$,
which behaves asymptotically like a decreasing exponential:
\begin{equation}\label{dndmtthermisch}
{{dn}\over{m_Tdm_T}} = {V\over{2\pi^2}} m_T K_1\Big({{m_T}\over{T}}\Big)
\; {\buildrel {m_T\gg T} \over \longrightarrow } \; V' \sqrt{m_T} \, e^{-m_T/T}
\end{equation}

\begin{figure}[htbp]
	\begin{minipage}[b]{\medfigsize}
		\epsfxsize \medfigsize \epsfbox{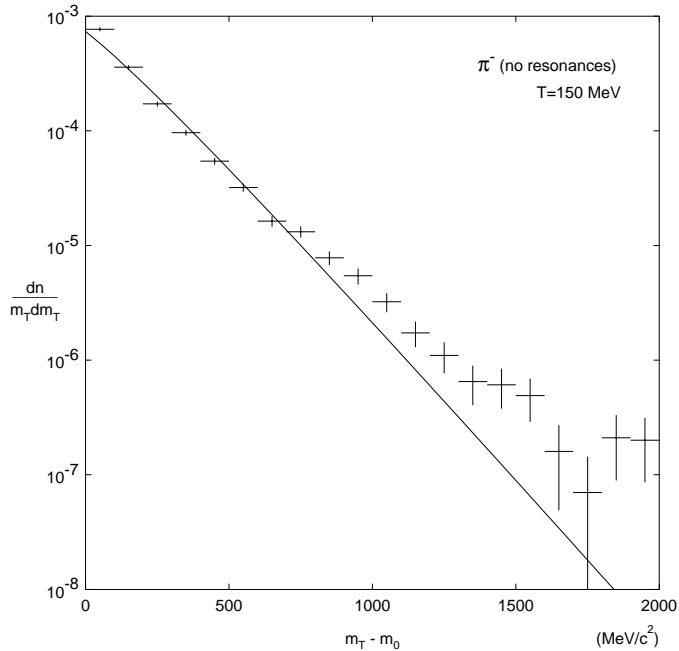}
	\end{minipage}
	\hfill
	\begin{minipage}[b]{\medcapsize}
		\caption[]{ \label{dmtpithnoreso} \sloppy
			{\bf \boldmath Purely thermal $\pi^-$ $m_T$-spectrum
			\unboldmath}
			in comparison with data from NA35 S+S 200 A GeV
			\cite{NA35charged}. A fit of a purely
			thermal source (eq. \ref{dndmtthermisch})
			was attempted to the data with the
			temperature $T$ as a free parameter. The absolute
			normalization was adjusted best possible,
			nevertheless the deviations at medium and high $m_T$
			are significant.
			}
		\vspace{0.5truecm}
	\end{minipage}
\end{figure}

While the basic characteristics of the measured $m_T$-spectra is indeed the
exponential decay over several orders of magnitude with
an almost uniform slope $1/T$ when plotted over $m_T$,
the $\pi^-$ spectrum from NA35
shows a significant concave curvature (Fig.~\ref{dmtpithnoreso})
on top of the exponential dropoff, which is not visible in
the $m_T$-spectra of the other particles ($p$, $K^0_s$, $\Lambda$,
$\overline \Lambda$). This curvature makes it impossible for a single thermal
source to fit both the low and the high $m_T$ region at the same time.
(It does not stem from the curvature
of the $K_1$ function, since that one is much smaller and in the
opposite direction.) The effect has been known under the name
{\it ``low-$m_T$ enhancement''} since the first data
from the ultrarelativistic heavy-ion experiments became available
\cite{otherdata},
and has stimulated many theoretical interpretations
(for an overview see e.g. \cite{Schukraft}).

Before we will devote ourselves to that problem in detail
we finish the discussion of the thermal source
with the rapidity distribution, which results from integrating the
invariant momentum spectrum over the transverse components:
\begin{equation} \label{dndythermal}
{{dn_{\rm th}}\over{dy}} = { V \over {(2\pi)^2} } T^3
	\Big(
	{{ m^2 }\over{T^2}}  + {m\over T} {2\over{\cosh y}}
				+ {{2}\over{\cosh^2 y}} \Big)
	\exp\left(-{m\over T} \cosh y \right) \; .
\end{equation}
which reduces for light particles to
$dn/dy \propto \cosh^{-2}(y-y_0)$ up to third order in $m/T$.
This rephrases the fact that the rapidity distribution of massless particles
from an isotropic source is always the same regardless
of their momentum dependence (e.g.~\cite{Gersdorffdndy}).
Its full width at half height $\Gamma^{\rm fwhm}_{\rm th} \approx 1.76$
is in sharp contrast to the experimental value for the pions of
$\Gamma^{\rm fwhm}_{\rm exp} = 3.3\pm 0.1$ \cite{NA35charged} (see dotted line
in Fig.~\ref{dypi}). The particle
mass further narrows the rapidity distributions, leading for $m > T$
to an additional gaussian with the width parameter $\sigma^2=m/T$,
and increases the deviations from the measured distributions.
Clearly also here are improvements necessary, which we will attempt
in section~\ref{flowingsources}.

\subsection{Resonance Decays}
\label{resonancedecays}

The attitude of the previous section was a bit too naive:
Apart from the directly emitted pions of assumedly thermal origin
the detector also detects pions, which originate from the decay of resonances,
which are also generated in the reaction, e.g.\
\begin{displaymath}
\rho^0 \rightarrow \pi^+ \pi^-
\;\hbox{,}\qquad
\omega \rightarrow \pi^+ \pi^0 \pi^-
\qquad\hbox{or}\qquad
\Delta \rightarrow N \pi^-
\end{displaymath}
and which have a very different spectral shape, thus distorting
the straight exponential dropoff.

The abundant production of these resonances is an experimental fact,
which has been investigated closely in $pp$ collisions \cite{ppreso}, where
because of the much lower multiplicities the individual resonances can be
reconstructed. Using these data as a guidance for our implementation in
nucleus--nucleus collisions, we find that the absolute multiplicities of the
resonances are distributed like $\propto \exp(-m_R/T)$ with
$T\approx 180\,\rm MeV$, justifying a thermal descripton as an approximation
as long as the temperature is close to this value. For the baryonic resonances
($\Delta,\Lambda^*,\ldots$) we also fix the baryon chemical potential to be
$\mu_B = 200\,\rm MeV$ and distribute the strange resonances
($K^*$, $\Lambda^*$, $\ldots$) according to strangeness
neutrality~\cite{strangeneut}. The spectral shape for the resonances
has also been shown in $pp$ experiments to be exponential with
$T\approx 180\,\rm MeV$ and can for the heavy-ion collisions quite safely be
assumed to be thermal since frequent elastic collisions will equilibrate the
resonances in the same way as the ground state particles.

The decays of the heavier resonances into their lighter daughter
particles has been investigated in detail by Sollfrank {\it et al.} in
\cite{Sollfrank} and others \cite{Barzreso}.
Since we will analyse the mechanism in detail
later in App.~\ref{slopesresonances},
we will here only sketch the method of computation.
The spectrum of the daughter particles is obtained by integrating
the momentum distribution of the resonances with a factor $f(W,p_R,p)$
describing the decay phase space \cite{Sollfrank}
within the kinematic limits of invariant mass
$W$, rapidity $y$ and transverse mass $m_T$ :
\begin{eqnarray}
\label{dymtdmtreso}
{{d^2n}\over{dy \,m_T dm_T}} &=&
	\int\limits_{W^{(-)2}}^{W^{(+)2}} dW^2
	\int\limits_{y_R^{(-)}}^{y_R^{(+)}} dy_R
	\int\limits_{m_{T R}^{2\,(-)}}^{m_{T R}^{2\,(+)}} dm_{T R}^2
	\,f(W,p_R,p)
	{{d^2n_R}\over{dy_R \,m_{T R} dm_{T R} }}
\; .
\end{eqnarray}
To compute the $\pi^-$ spectrum we evaluate (\ref{dymtdmtreso})
for the two-body decays
$\rho$, ${K^-}^*$, ${K^0}^*$, $\Delta$, ${\Sigma^-}^*$,
${\Sigma^0}^*$, $\Lambda^*$ as well as for the three-body decays of
$\omega$ and $\eta$ numerically.

\begin{figure}[tbp]
	\begin{minipage}[b]{\bigfigsize}
		\epsfxsize \bigfigsize \epsfbox{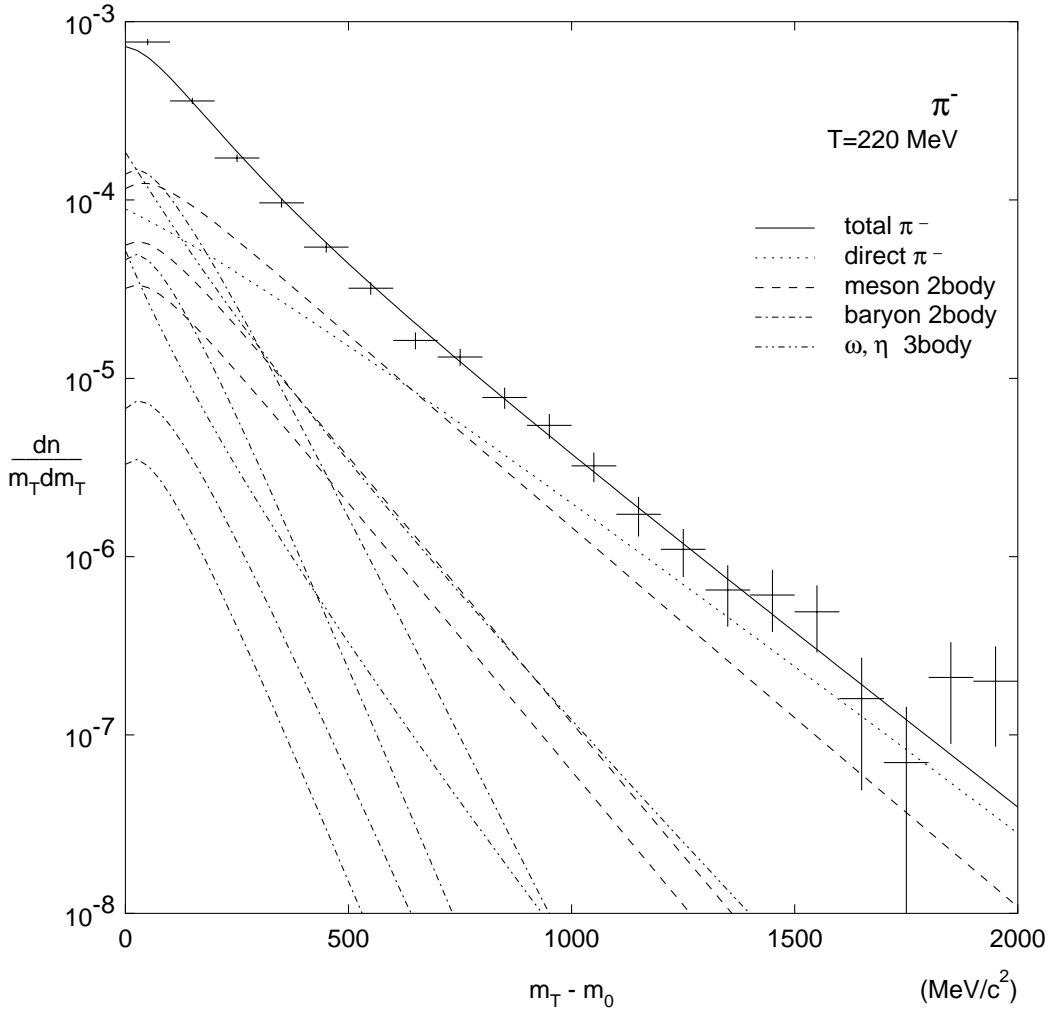}
	\end{minipage}
	\hfill
	\smallskip
	\hspace*{1 truecm}
	\begin{minipage}[b]{\bigcapsize}
		\caption[]{ \label{dmtpith} \sloppy
			{\bf \boldmath Thermal $\pi^-$ $m_T$-spectrum
			with resonance decays \unboldmath }
			in comparison to data from NA35
			$S$+$S$ $200\,A\,\rm GeV$ \cite{NA35charged}.
			The purely thermal $\pi^-$ and the two- and three-body
			decays of mesons and baryons add up to the total
			spectrum. The fit of the total $\pi^-$-spectrum
			from a thermal source with the temperature $T$
			as a free parameter is very good over the full range.
			The absolute normalization have been adjusted for a
			best fit, $\mu_b = 200 \rm\,MeV$ has been fixed.
			These results from~\cite{Sollfrank} are repeated
			here for later comparison within a coherent
			presentation.
			}
	\end{minipage}
\end{figure}

\begin{figure}[tbp]
	\begin{minipage}[b]{\bigfigsize}
		\epsfxsize \bigfigsize \epsfbox{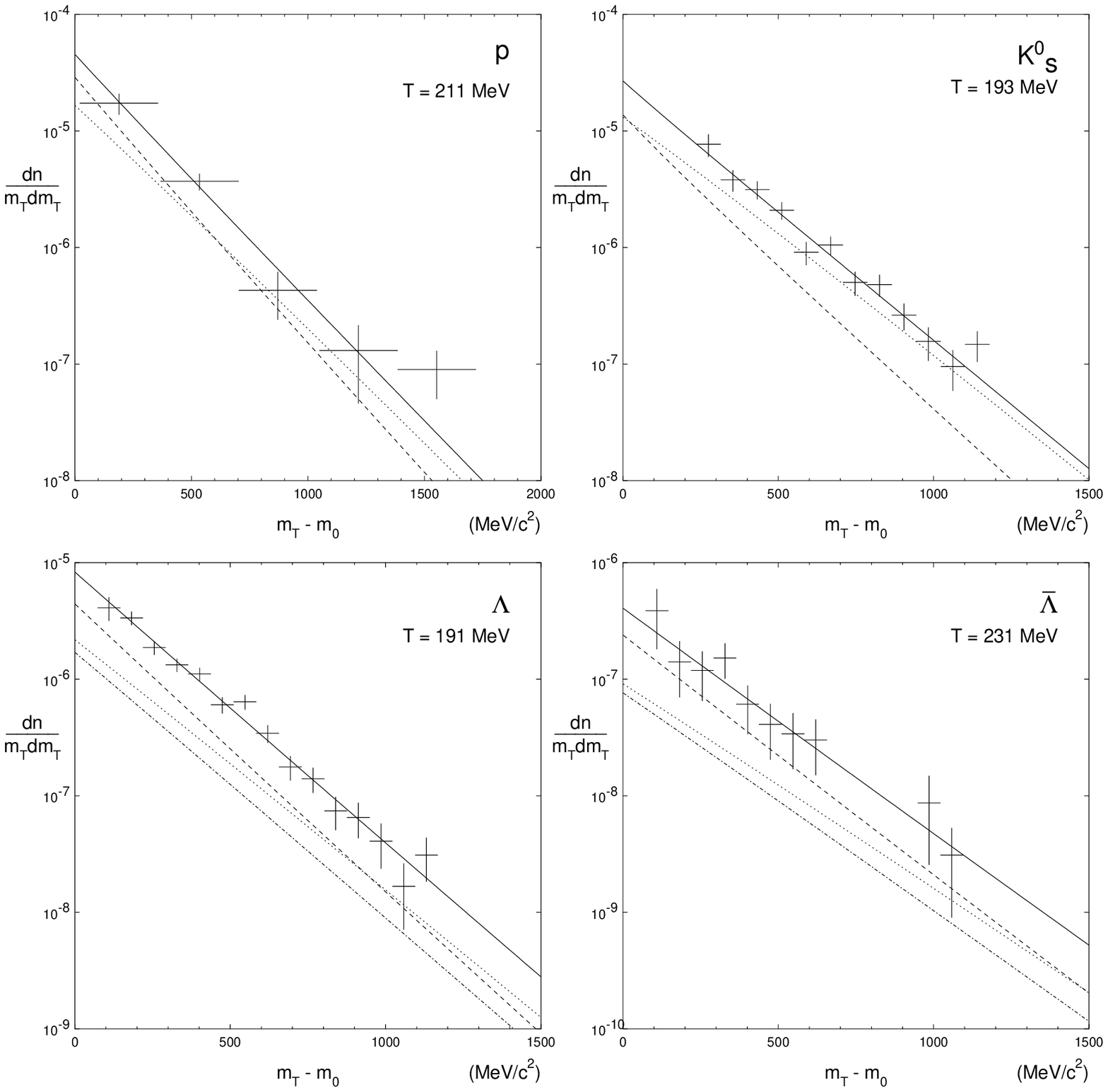}
	\end{minipage}
	\smallskip
	\hspace*{0.7 truecm}
	\begin{minipage}[b]{\bigcapsize}
		\caption[]{ \label{dmtallth} \sloppy
			{\bf \boldmath Thermal $m_T$-spectra of $p$, $K^0_s$,
			$\Lambda$ and $\overline \Lambda$
			including resonance decays \unboldmath}
			in comparison with data from NA35 $S$+$S$ 200
			GeV/n \cite{NA35charged,NA35strange}.
			The purely thermal spectra (dotted) and the
			resonance decays (dashed) add up to the total spectra
			(solid). The respective fits to the $m_T$-spectra
			are all very good and give similar temperatures.
			The absolute normalizations have been adjusted for a
			best fit,  $\mu_b = 200 \rm\,MeV$ has been fixed.
			These results from from~\cite{Sollfrank} are repeated
			here for later comparison within a coherent
			presentation.
			}
	\end{minipage}
\end{figure}

With these assumptions the transverse mass spectrum of the pions is
again only a function of the temperature $T$ and baryon chemical potential
$\mu_b$ and one can try a fit to the
data. As shown in Fig.~\ref{dmtpith}, we now succeed in reproducing
the spectrum over the whole range in $m_T$, with the temperature $T$
corresponding to the slope at high $m_T$. The kinematics of the resonance
decays result in very steeply dropping daughter pion spectra
and raise considerably the total pion yield at low $m_T$.
For the heavier  $K^0_s$, $p$, $\Lambda$ and
$\overline \Lambda$ (Fig.~\ref{dmtallth}) the change in the spectra
is less pronounced, since the mass difference between the resonance and the
ground state is much smaller. Quite remarkably, we realize that
all independent fits to these five particle spectra seem
to agree in $T\approx 200\rm\, MeV$
within their statistical uncertainties~\cite{Stroebele,Ornikreso}.

We conclude that the resonance decays, as a necessary ingredient
for any thermal model, provide already
a good description of all measured hadronic $m_T$-spectra.
The magnitude of the effect (2/3 of the pions come from resonances)
as it was assumed here, is caused by the high fit temperature and
has not been confirmed by further data, e.g.\ on the relative multiplicities
of $\rho$, $\omega$, $\Delta$, etc.\ in $S$+$S$, so that there is still
some room for other explanations (which should, however, always also take
the resonances into account, albeit perhaps at a somewhat reduced level).

Especially the possibility of a chemical potential
for the pions in size of $\mu_\pi \approx 118\rm\,MeV$ at $T=164\rm\,MeV$
has been discussed
\cite{pionpotential}, which might originate from non-equilibrium effects and
which would accumulate pions at low momenta
because of the nearby divergence of the Bose distribution.
With our treatment of the $\pi^-$-spectra we cannot exclude
this effect, but we can reduce its importance by first subtracting the
always present resonance contributions.
In a quantitative analysis of the pion multiplicity in our hydrodynamic model
\cite{hydrodraft}, we determine that the
non-equilibrium population of pions and the accompanying pion potential
is much smaller than the pion mass and cannot influence
the shape of the $m_T$-spectra via the upwards curvature of the
Bose distribution.

For a further overview of the low $m_T$-enhancement we would
like to refer to \cite{Schukraft}. From our side
the possibility of transverse flow was brought into the
discussion with a model based on spherical
symmetry \cite{Search}.
Here, however, we will adopt cylindrical symmetry and
analyse the longitudinal and transverse flows separately.
While the longitudinal flow can be extracted from the data,
the same uniqueness cannot be achieved for the transverse flow,
though consistency arguments suggest its existence.

\section{Spectra From Flowing Sources}
\label{flowingsources}

\subsection{Longitudinal Flow}
\label{longitudinalflow}

We have already mentioned that the rapidity distribution of $\pi^-$
(actually negatively charged hadrons) as measured by NA35
is much wider than a stationary thermal source could possibly predict.
Also the inclusion of resonance decays does not improve this picture,
since in the same way as they lead to a steepening of the $m_T$-spectra
at low $m_T$ they cause a slight narrowing of the $y$-distribution.

Obviously the momentum distribution of the measured pions is not isotropic,
it rather has imprinted on it the direction of the colliding nuclei.
The {\it boost-invariant} longitudinal expansion model,
as it has been postulated by Bjorken
\cite{Bjorken} can explain such an anisotropy already at the level of
particle production, which subsequently, after sufficient rescattering,
leads to a boost-invariant longitudinal flow of matter
with locally thermalized distributions.
The observed particle spectrum then results from
the summation of the spectra of
individual thermal sources which are uniformly distributed in flow angle
$\eta$. However, since the model
has been formulated for asymptotically high energies, where the rapidity
distribution
of the produced particles establishes a plateau at midrapidity,
we cannot apply it directly to our current energies, where
already half of the total rapidity gap of 6 units is eaten up by
the target and projectile fragmentation regions and consequently the pion
distribution rather looks like a gaussian.

We modify the boost invariant scenario to account for the
limited available beam energy by restricting the
boost angle $\eta$ to the interval $(\eta_{\rm min}, \eta_{\rm max})$
\cite{Sinyukovflow}.
The rapidity distribution is then the
integral over the uniformly distributed thermal sources
(\ref{dndythermal}) boosted individually by $\eta$:
\begin{equation}\label{dndythint}
{{dn}\over{dy}}(y) =
	\int\limits_{\eta_{\rm min}}^{\eta_{\rm max}} \! d\eta \,
		{{dn_{\rm th}}\over{dy}} (y-\eta)
\end{equation}
The transverse mass spectrum is not affected by this operation.

\begin{figure}[p]
	\begin{minipage}[b]{\bigfigsize}
		\epsfxsize \bigfigsize \epsfbox{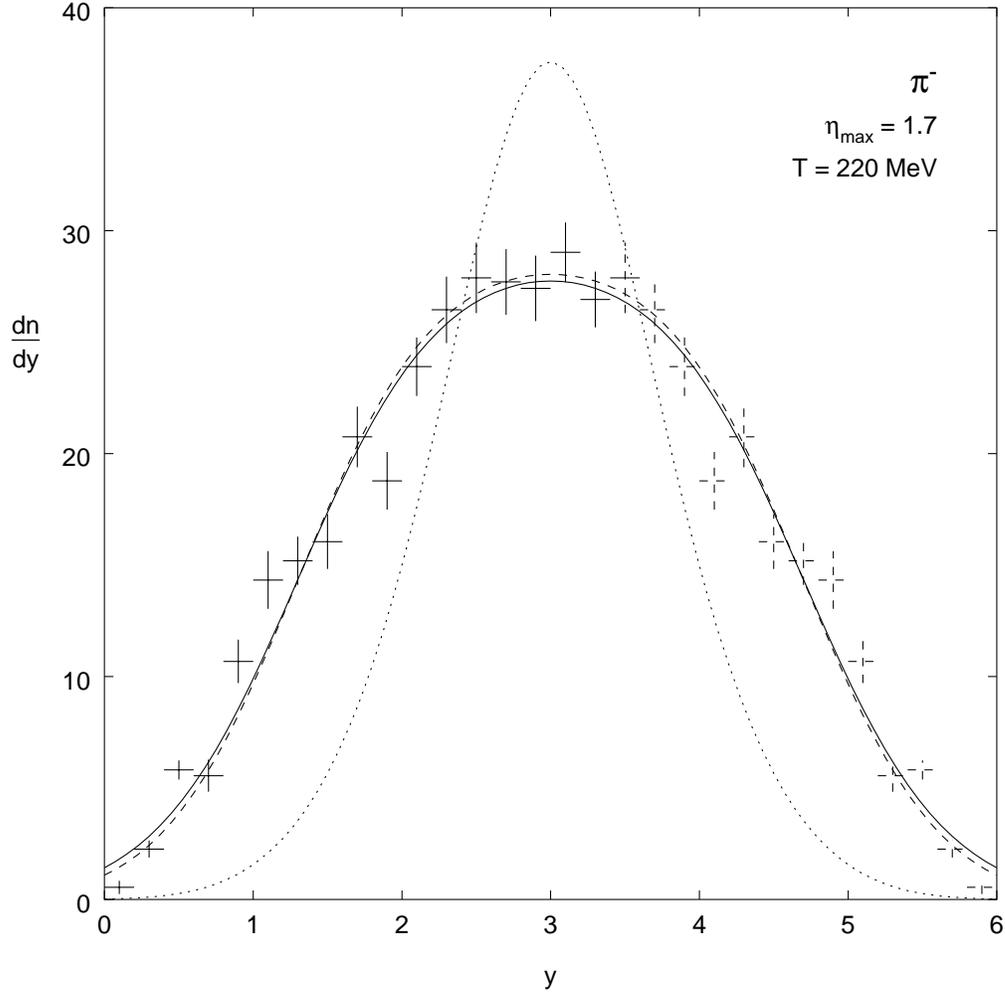}
	\end{minipage}
	\smallskip
	\hspace*{1 truecm}
	\begin{minipage}[b]{\bigcapsize}
		\caption[]{  \label{dypi} \sloppy
			{\bf \boldmath $\pi^-$ $y$-spectrum with longitudinal
			flow \unboldmath}
			in comparison with data from NA35 $S$+$S$
			$200\,A\,\rm GeV$ \cite{NA35charged}, which
			have been reflected around $y=3$.
			The maximal fluid rapidity $\eta_{\rm max}=1.7$
			has been obtained from a fit of eq.~(\ref{dndythint})
			to the measured spectrum. The absolute
			normalization has been adjusted for a best fit,
			the agreement with the data is very good.
			The temperature $T=220\,\rm MeV$ has been taken from
			the fit to the $m_T$-spectrum and has only a
			negligible influence on the $y$-spectrum,
			similar to the inclusion of resonance decays
			(dashed line).
			For comparison also the purely thermal spectrum
			without any longitudinal flow is shown as a dotted
			curve.
			}
	\end{minipage}
\end{figure}

\begin{figure}[p]
	\begin{minipage}[b]{\bigfigsize}
		\epsfxsize \bigfigsize \epsfbox{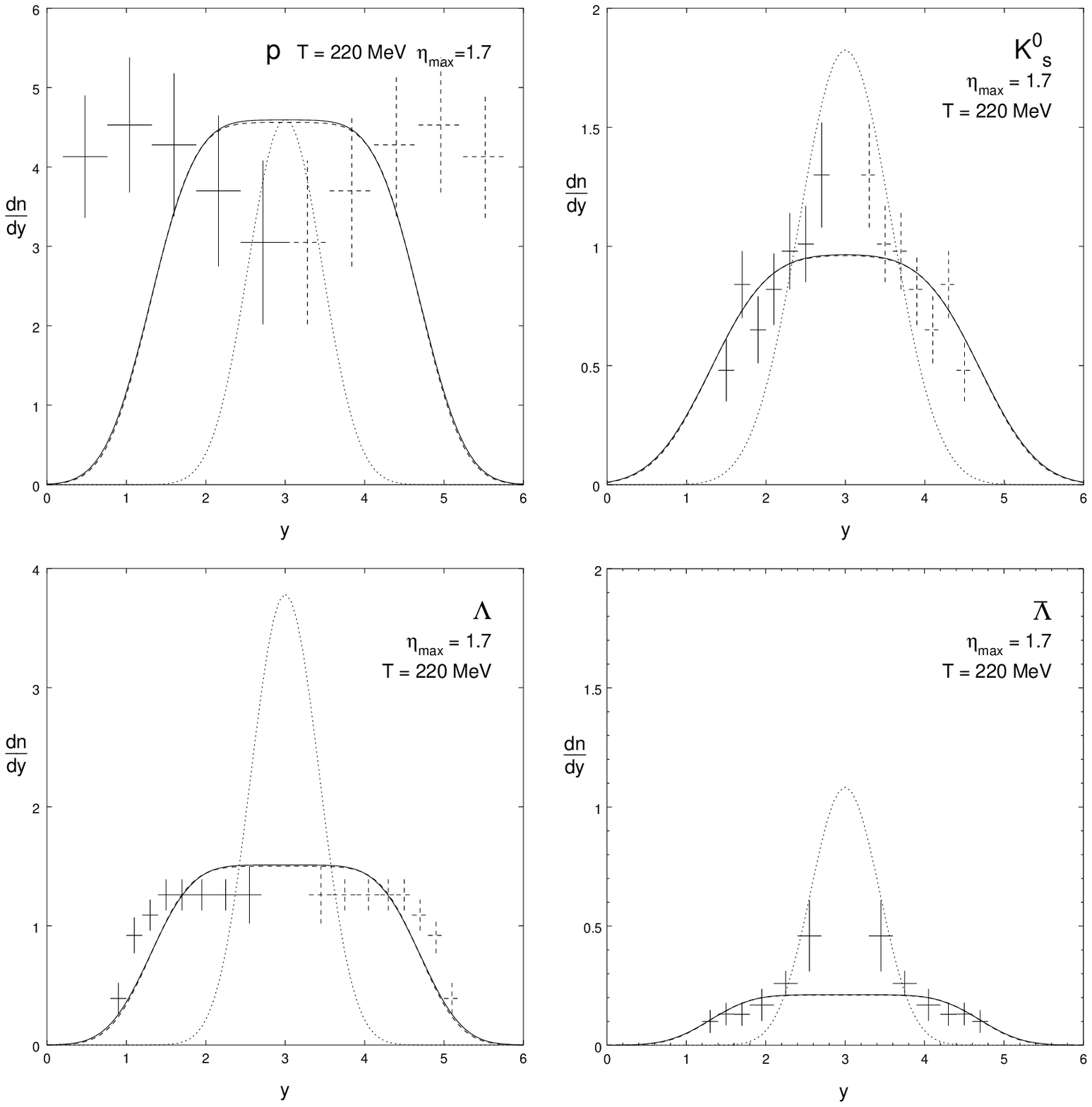}
	\end{minipage}
	\smallskip
	\hspace*{0.7 truecm}
	\begin{minipage}[b]{\bigcapsize}
		\caption[]{ \label{dyall} \sloppy
			{\bf \boldmath $y$-spectra of $p$, $K^0_s$,
			$\Lambda$ and $\overline \Lambda$
			with longitudinal flow\unboldmath}
			in comparison with data from NA35 $S$+$S$
			$200\,A\,\rm GeV$ \cite{NA35charged,NA35strange}.
			The flow $\eta_{\rm max} =1.7$ seen in the pions
			(Fig.~\ref{dypi})
			describes all produced spectra satisfactorily, only
			the protons still carry a large fraction of their
			initial collision energy.
			The absolute normalization was always adjusted for a
			best fit, the temperature and the resonance decays
			(dashed) have no influence on the spectra.
			For comparison also the purely thermal spectra
			without longitudinal flow are shown as dotted curves.
			}
	\end{minipage}
\end{figure}

We can now try a fit of (\ref{dndythint}) to the measured $\pi^-$ $y$-spectrum
using $\eta_{\rm max}=-\eta_{\rm min}$ as the single free parameter,
because the distributions are not very sensitive to $T$. The rapidity
distribution of
negative hadrons from NA35 is measured over a large acceptance region
($0<p_T<2{\,\rm GeV}/c$) and can be identified with the $\pi^-$, since
the additional contributions from $K^-$ and $\bar p$ are small (5-10\%)
and probably distributed more or less homogeneously over the whole region
\cite{NA35charged}.
The data points in the not accessible region of the streamer chamber
above $y>3.4$ can be substituted using symmetry with respect to $y=3$.

We see in Fig.~\ref{dypi} that $\eta_{\rm max} = 1.7$ fits the measured
$\pi^-$-spectra quite nicely. Again, the inclusion of the resonance decays
does not qualitatively change the $\pi^-$ distribution since the resonances are
uniformly
distributed over the boost rapidity $\eta$ and have thus the same minor effect
on the integral (\ref{dndythint}) as they have on the individual thermal
spectra (\ref{dndythermal}). This would be different for inhomogeneities in the
resonance distributions: for example a high concentration of $\Delta$'s
in the fragmentation regions would manifest itself by additional pions
from that region, i.e.\ bumps in the rapidity distribution \cite{Ornikreso}.

Looking at the other hadronic distributions (Fig.~\ref{dyall}), we see our
interpretation basically confirmed. However, the experimental proton
spectrum is much broader than the computed one and has a dip
at central rapidity; the measured protons
obviously carry still a big amount (50\%) of their initial collision energy
\cite{Stroebele}, and the spectrum contains many protons near the
target and projectile fragmentation regions, which have not suffered
sufficiently many collisions to have become part of the central fireball.
Because there is no clearcut separation
between the central region and the fragmentation regions in the
experimental $y$-spectrum, our model, which is crafted for
the central region only, is bound to fail for the fragmentation zone protons.

The produced strange particles are much better reproduced:
the theoretical $K^0_s$ spectrum is only slightly too broad,
the one for $\Lambda$'s slightly
too narrow and the one for $\overline \Lambda$'s still satisfactory with the
exception
of the central data point which also has the largest statistical uncertainty.
The strange particles are good indicators
for collective flow, because all are produced particles which were absent
in the original nuclei and are thus not contaminated by a cold spectator
component at leading rapidities as the protons are. Also because of their
bigger mass their flow component is more accentuated
against the random thermal motion as in case of the pions.
Moreover, the disappearance of the characteristic polarization
of the $\Lambda$ \cite{NA35strange} hints towards at least one
other collision in the medium, which justifies viewing them as part of a
thermalized system. For the $\overline \Lambda$ the situation is more
complicated and the success of our model might be only accidental.
We can imagine that they are produced centrally in
particularly hard individual nucleon--nucleon collisions and
have not yet approached thermal and chemical equilibrium between annihilation
and production. We could thus expect to see in the $\overline \Lambda$
distribution besides the flow component the imprint of the production process
at central rapidities. However, a recent chemical analysis of the various
particle ratios \cite{LetessierTounsi}, which views the $\overline\Lambda$'s
as an integral part of a thermally and chemically equilibrated fireball
appears to be phenomenologically very successful and in favor of a
thermalized picture which has already lost its memory of the production
process.

\begin{figure}[tb]
	\begin{minipage}[b]{\smallfigsize}
		\epsfxsize \smallfigsize \epsfbox{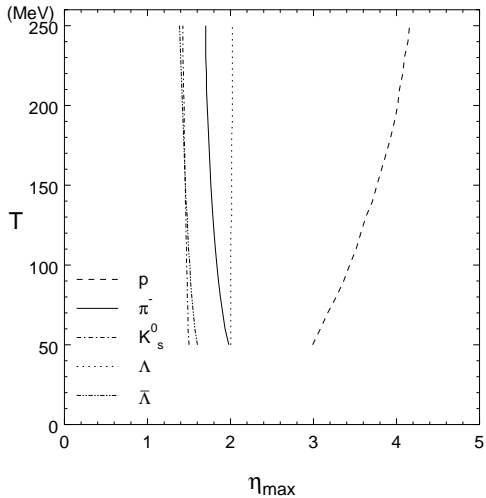}
	\end{minipage}
	\hfill
	\begin{minipage}[b]{\smallcapsize}
		\caption[]{ \label{fityall} \sloppy
			{\bf \boldmath The longitudinal flow $\eta_{\rm max}$
			\unboldmath }
			can be extracted from the rapidity distributions
			of the hadrons. With exception of the protons,
			which carry still a large fraction of their initial
			beam motion, a fit with $\eta_{\rm max}$ as a free parameter
			gives for all produced particles independent of the
			assumed temperature a longitudinal flow
			similar to the pions with $\eta_{\rm max}\approx 1.7$
			at $T\approx 200 \,\rm MeV$.
			}
		\vspace{1 truecm}
	\end{minipage}
\end{figure}

We can also try a best fit of the hadronic $y$-spectra individually
by adjusting $\eta_{\rm max}$ for each particle species separately
(Fig.~\ref{fityall}). Again we arrive at the same value of
$\eta_{\rm max} = 1.7 \pm 0.3$ independent of the temperature. Unfortunately,
the $y$-distributions of the heavier particles have less statistics
than the pion spectrum and do not extend to as small rapidities;
hence the estimates of the widths relies heavily on the
last data points. Therefore we will prefer to continue our reasoning
with the better determined pion value.

\subsection{Transverse Flow}
\label{transverseflow}

The stepwise extension of the initial stationary thermal model by
including resonance decays and a longitudinal flow component
has provided a phenomenological description
of almost all spectra -- $dn/dy$ as well as $dn/dm_T^2$ --  of pions,
protons, $K^0_s$, $\Lambda$ and $\overline \Lambda$ as measured
in the experiment NA35. The few exceptions can be explained, they show the
limits of our model. However, this seemingly complete picture of the dynamics
of the hadronic matter has some serious intrinsic theoretical problems:

\begin{enumerate}

\item
On one hand at temperatures around $200\,\rm MeV$ the mean free path
of pions in hadronic matter is much less than $1 \rm\, fm$,
as we can estimate from the thermal distributions and
averaged cross section of pions with nucleons and with each other
\cite{Sollfrank}.
On the other hand the size and lifetime of the reaction region is
several ${\rm fm}$, as can be guessed from the radius of the $^{32}S$
nucleus and causality. The length of the collision zone will be even larger
because the large time dilatation factors of the relativistic
longitudinal expansion stretch the system quite long already
during the formation time $\tau_f\approx 1 {\,\rm fm}/c$.
Consequently, the pions cannot leave the interaction zone
at $T\approx 200 \,\rm MeV$ without further collisions,
the reaction region cannot decouple thermally
and should by continuing expansion force the pions to cool down further.

\item
The longitudinal expansion in the boost-invariant scenario corresponds to
a solution of the hydrodynamical equations in (1+1)-dimensions.
In fact the 1-dimensional expansion presumably dominates
initially because of the anisotropic initial conditions.
But it is inconsistent to assume that a thermalized system
expands collectively in longitudinal direction
without generating also transverse flow from the high pressures
in the hydrodynamic system.
This should be computed by hydrodynamics in at least (2+1) dimensions,
if azimuthal symmetry is maintained, otherwise (3+1)-dimensions,
because the transverse expansion will increase the cooling rate.

\item
Already transverse velocities of about $0.3$-$0.6\,c$, which are moderate
compared to the longitudinal expansion with
$\beta_L=\tanh \eta_{\rm max} > 0.9 \,c$,
could change the transverse mass spectra considerably by
enhancing the particle yields at high $m_T$.

\end{enumerate}

We want to address these problems with a model which contains
resonance decays and both longitudinal and transverse flow in order to
put the discussion of the longitudinal and transverse spectra
on a common basis. In this paper we will analyse how much information
can be extracted from the data with such a model.
Using our phenomenological analysis presented here we
will extend in the next step \cite{circumstantial,hydrodraft}
the model towards a global hydrodynamical description which enforces
the mentioned theoretical consistency requirements.
We observe (similar to \cite{Ornikreso}) that agreement
with all the presented data can again be achieved and we prove the existence
of transverse flow by theoretical means \cite{circumstantial,hydrodraft}.

\begin{figure}[tbp]
	\begin{minipage}[b]{\bigfigsize}
		\epsfxsize \bigfigsize \epsfbox{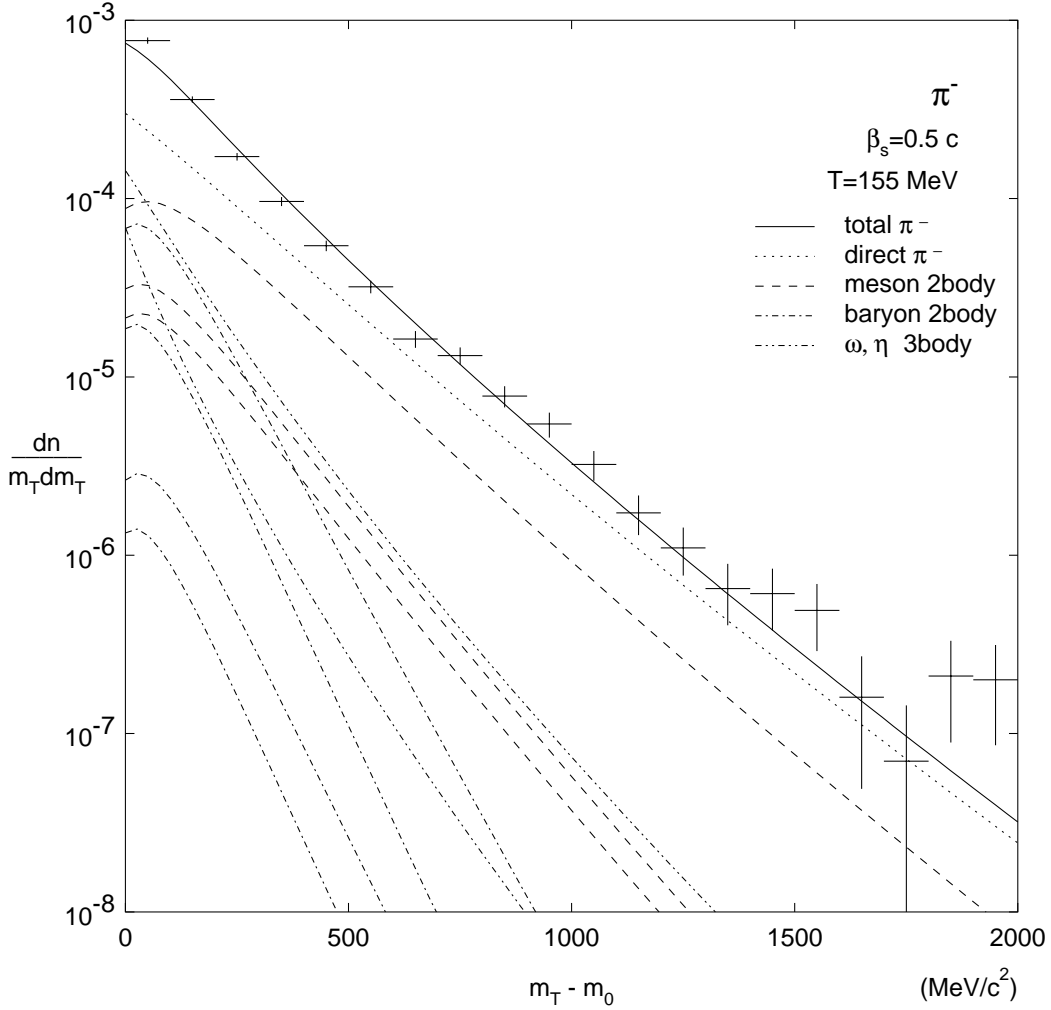}
	\end{minipage}
	\smallskip
	\hspace*{1 truecm}
	\begin{minipage}[b]{\bigcapsize}
		\caption[]{ \label{dmtpi2} \sloppy
			{\bf \boldmath $\pi^-$ $m_T$-spectrum
			with resonance decays and transverse flow
			\unboldmath }
			in comparison with data from NA35 $S$+$S$
			$200\,A\,\rm GeV$ \cite{NA35charged}.
			The total $\pi^-$ spectrum
			consists of the original pions and the decay pions.
			Based on a parabolic transverse velocity profile with
			$\beta_s=0.5\, c$ the fit with the temperature $T$
			as the free parameter is quite good over the whole
			$m_T$-range.
			The absolute normalization has been adjusted for
			a best fit, $\mu_b = 200 \rm\,MeV$ has been fixed
			to allow for a moderate baryon population.
			}
	\end{minipage}
\end{figure}

\begin{figure}[tbp]
	\begin{minipage}[b]{\bigfigsize}
		\epsfxsize \bigfigsize \epsfbox{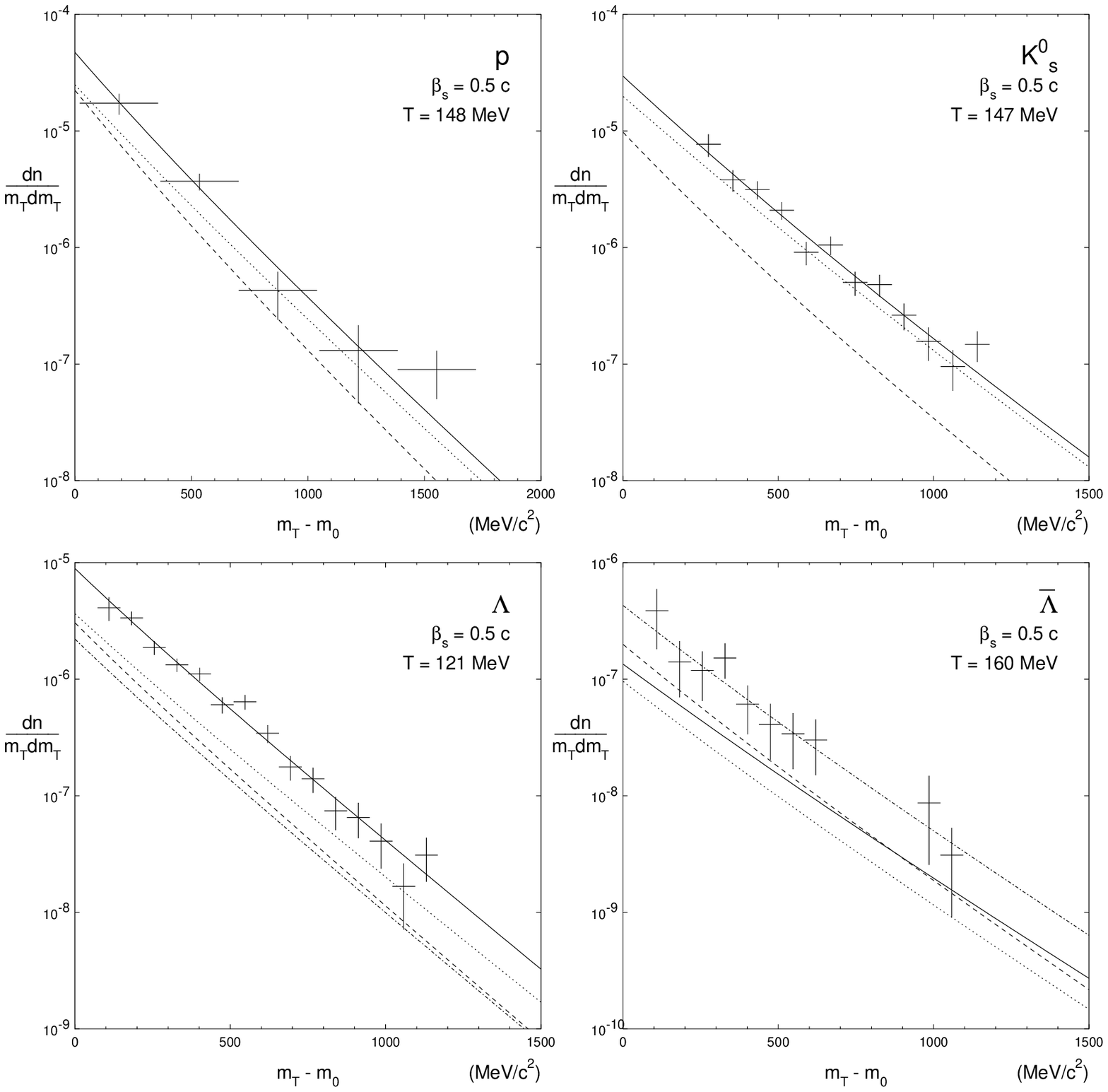}
	\end{minipage}
	\smallskip
	\hspace*{0.7 truecm}
	\begin{minipage}[b]{\bigcapsize}
		\caption[]{ \label{dmtall2} \sloppy
			{\bf \boldmath The $m_T$-spectra of $p$, $K^0_s$,
			$\Lambda$ and $\overline \Lambda$ with decays and
			transverse flow \unboldmath}
			in comparison with the data from NA35 $S$+$S$ at
			$200\,A\,\rm GeV$~\cite{NA35charged,NA35strange}.
			Based on a parabolic transverse velocity profile with
			$\beta_s=0.5\,c$ the individual fits
			with the temperature as the free parameter show
			good agreement with the data at temperatures
			similar to Fig.~\ref{dmtpi2}.
			The absolute normalization has been adjusted for
			a best fit, $\mu_b = 200 \rm\,MeV$ has been fixed
			as in Fig.~\ref{dmtpi2}.
			}
	\end{minipage}
\end{figure}

Not only theoretical necessities suggest the inclusion of transverse flow.
There are further arguments for the actual existence of the phenomenon:
\begin{enumerate}

\item
At the lower BEVALAC energies ($E/A \le 2\rm\, GeV$) a sidewards flow
of nuclear matter is observed~\cite{BEVALAC}. It is generated by the
squeeze-out of nuclear matter in the collision and has been predicted by
hydrodynamic computations~\cite{Stoecker}. Though this directed flow
is of different nature as our collective expansion flow,
it suggests the relevance of hydrodynamics also at higher energies.

\item
In a simple picture the transverse flow flattens, in the region
$p_\perp \alt m$,
the transverse mass spectra of the
heavier particles more than for the lighter
particles~\cite{SiemensRasmussen,Search}.
The AGS data~\cite{E802} show clearly this tendency for pions, kaons and
protons.

\item
Event generators for nucleus--nucleus collisions show a statistical effect
which can be interpreted as transverse flow. The statistical averaging
of the velocities of many particles in a small volume can exhibit
the transverse flow at AGS energies \cite{RQMDflow}.
The VENUS event generator for nucleus--nucleus collisions
shows also for CERN-energies an increase of the spectra at high $m_T$,
once the produced particles are allowed to rescatter~\cite{Venus}.
The same effect is produced in our model by the transverse flow.

\end{enumerate}

On the other hand we have to admit that from phenomenology alone
there is no need for an additional transverse flow effect,
since all the data have already
been described nicely. Currently we thus will not be able to prove
the existence of transverse flow from the data alone, only an
indirect proof based on the phenomenological analysis together with
hydrodynamical calculations with a consistent theoretical treatment
of the freeze-out process can be given \cite{circumstantial}.
But in the near future bigger collision systems can be expected
with the $Au$-beam at the AGS and with the installation of
the $Pb$-injector at CERN, which will lead to an
increase of the collective effects. If we are currently not able to prove
transverse flow directly from the data, the chances
will be improving tomorrow, and we should develop our tools in the meantime.

We compute the spectrum by boosting the thermal sources now both in
longitudinal
and transverse direction. We describe the transverse velocity distribution
$\beta_r(r)$ in the region $0\le r \le R$ by a self-similar profile, which is
parametrized by the surface velocity $\beta_s$
\begin{equation}
\label{betar}
\beta_r( r ) = \beta_s \left( {r\over R} \right)^n  \quad.
\end{equation}
With $n$ we can vary the form of the profile. We choose customarily
$n=2$, because the quadratic profile resembles the solutions of hydrodynamics
closest \cite{globalhydro}. Anyway, the form of the profile
is not important for the analysis, as we checked with the case $n=1$.
Leaving the details of the computation to App.~\ref{slopestransverseflow},
the resulting spectrum is a superposition of the individual thermal
components, each boosted with the boost angle $\rho = \tanh^{-1} \beta_r$:
\begin{equation}
{{dn}\over{m_T\,dm_T}} \propto
	\int_0^R r\,dr \, m_T 	I_0\Big(\frac{p_T \sinh\rho}{T}\Big)
				K_1\Big(\frac{m_T \cosh\rho}{T}\Big)
\end{equation}

{}From Figs.~\ref{dmtpi2} and \ref{dmtall2} we see that also with
a moderate transverse flow
($\beta_s=0.5\,c$, i.e. $\langle\beta_r\rangle=0.25\,c$) very good fits to all
hadron spectra can be obtained,
again treating the temperature as a free parameter. The temperature of the
local
sources as extracted from the data is again suprisingly uniform
but now much smaller than without transverse flow
(Figs.~\ref{dmtpith} and \ref{dmtallth}), thereby reducing
the contribution from the resonances.
A quantitative explanation of this effect  will be given later
in App.~\ref{slopestransverseflow},
qualitatively the effect is similar to a blue shift
as caused by a rapidly approaching source,
shifting particles to higher momenta. The heavier the particles, the more
they profit from the flow velocity.

\begin{figure}[tbh]
	\begin{minipage}[b]{\medfigsize}
		\epsfxsize \medfigsize \epsfbox{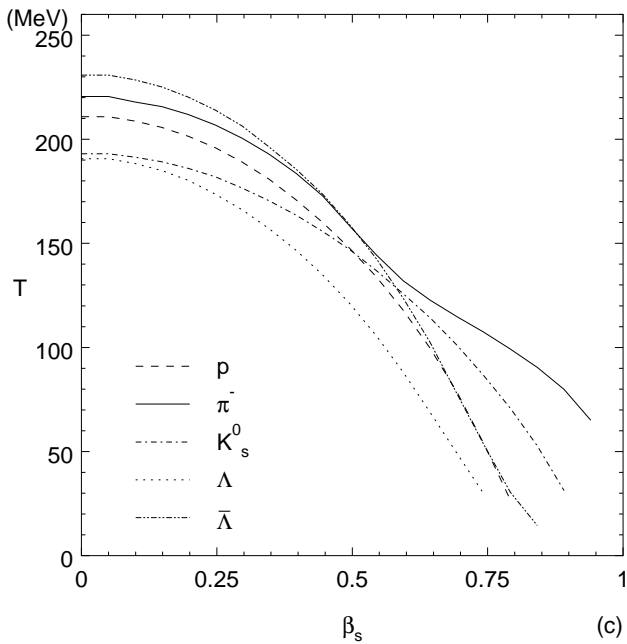}
	\end{minipage}
	\hfill
	\begin{minipage}[b]{\medcapsize}
		\caption[]{ \label{fit2all} \sloppy
			{\bf \boldmath All fitpairs  $(T,\beta_s)$
			of the $\pi^-$, $K^0_s$, $p$, $\Lambda$ and
			$\overline \Lambda$ $m_T$-spectra. \unboldmath}
			Every point on these curves gives a good agreement
			of the computed $m_T$ spectrum (with resonances
			and transverse flow) with the respective measured
			spectrum.
			}
		\vspace{0.5truecm}
	\end{minipage}
\end{figure}

\begin{figure}[htb]
	\begin{minipage}[t]{\twofigsize}
		\epsfxsize \twofigsize \epsfbox{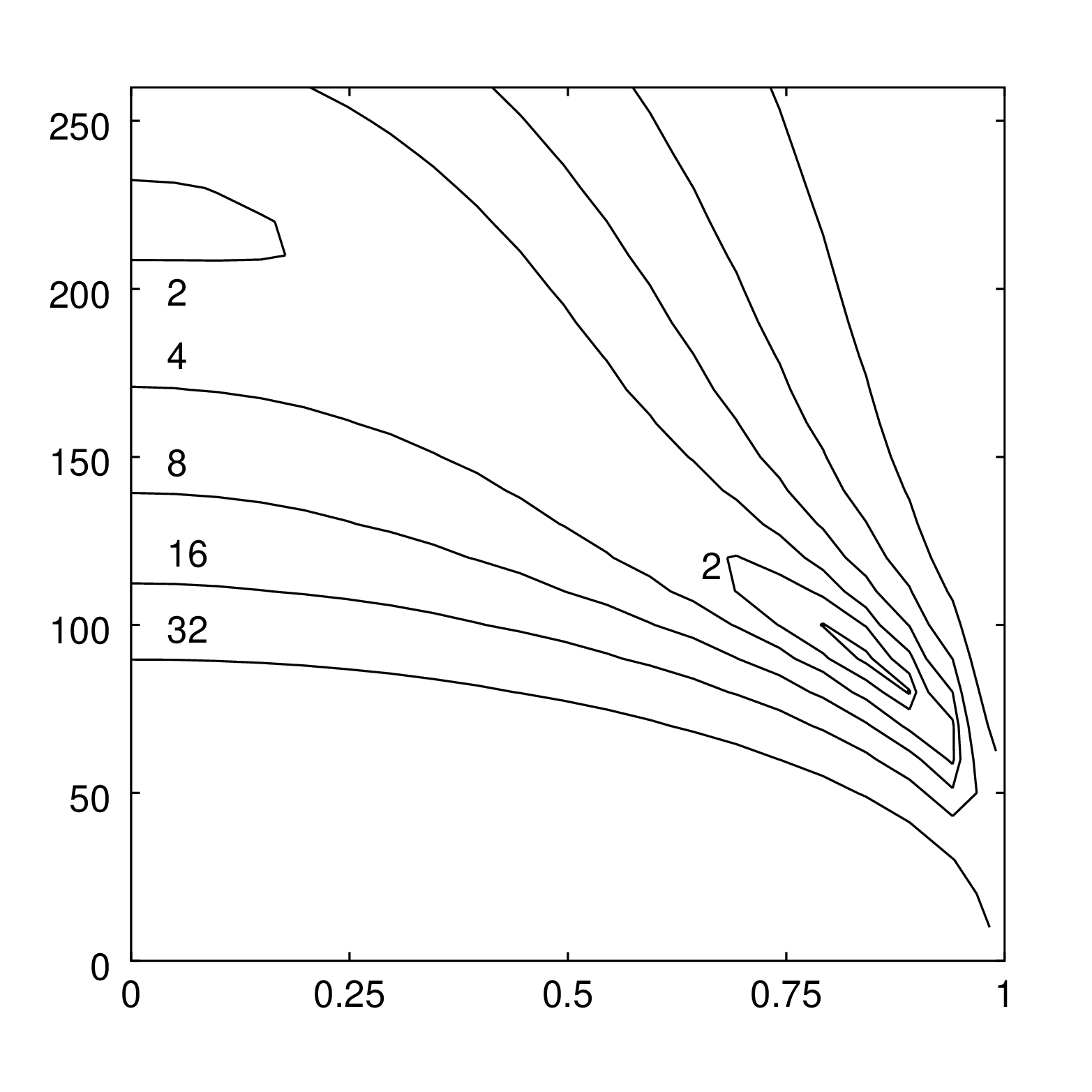}
		\begin{minipage}{\twofigsize}
		\hspace*{0.5truecm}
		\begin{minipage}{\twocapsize}
		\caption[]{\label{scan2pi} \sloppy
			{\bf \boldmath Quality of the $(T,\beta_s)$ fits
			for the $\pi^-$ $m_T$ spectra. \unboldmath}
			The contour lines designate the areas of
			$\chi^2/NDF = 1$, 2, 4, 8, 16, and 32.
			$\mu_b$ has been fixed at 200 MeV.
			}
		\end{minipage}
		\end{minipage}
	\end{minipage}
	\hfill
	\begin{minipage}[t]{\twofigsize}
		\epsfxsize \twofigsize \epsfbox{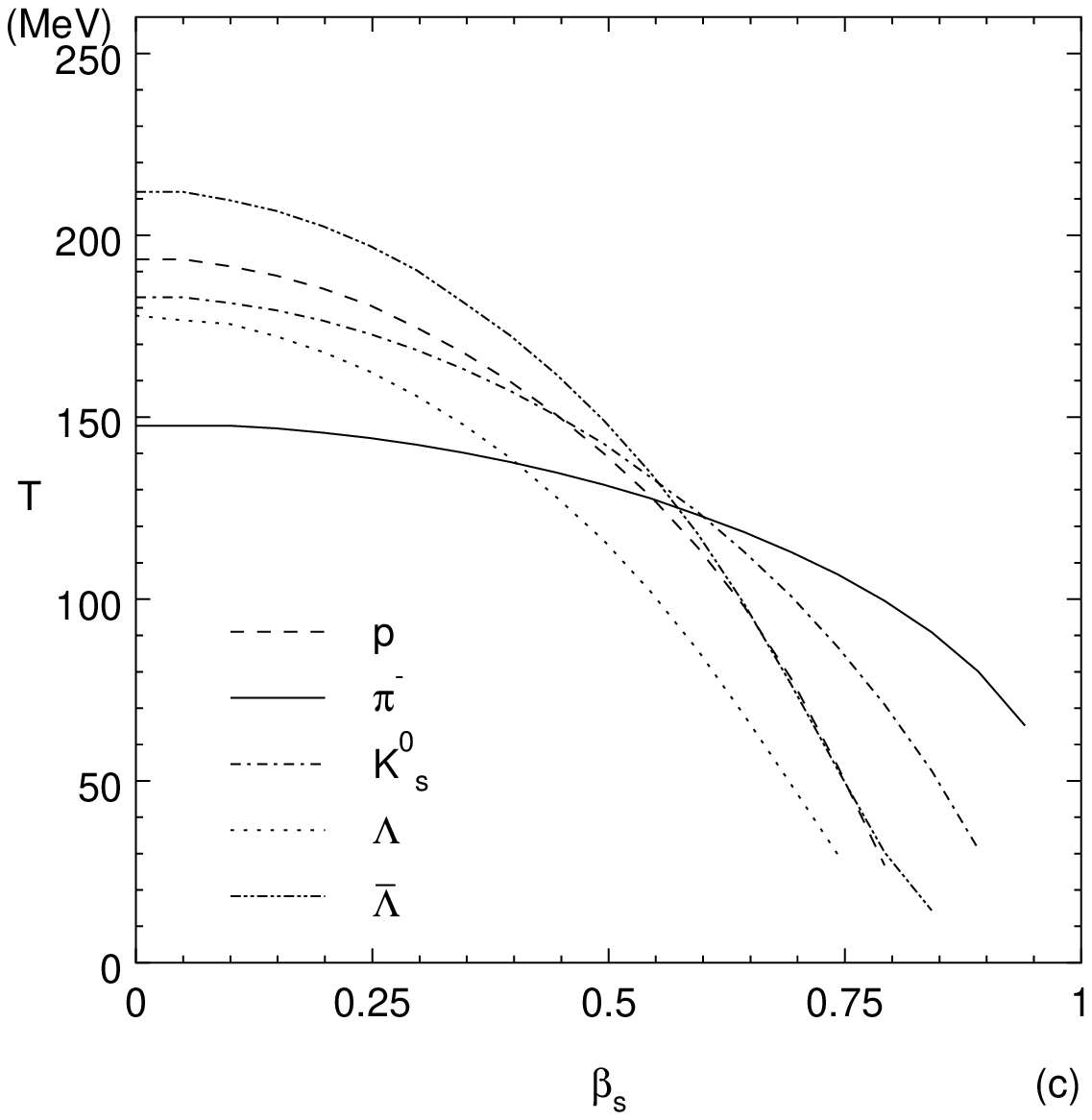}
		\begin{minipage}{\twofigsize}
		\hspace*{0.5truecm}
		\begin{minipage}{\twocapsize}
		\caption[]{\label{fit2allnoreso} \sloppy
			{\bf \boldmath Fit pairs of $\pi^-$, $K^0_s$, $p$,
			$\Lambda$ and $\overline \Lambda$
			$m_T$ spectra omitting the decay contributions.
			\unboldmath }
			In this unrealistic situation there would be an
			intersection around $\beta_s\approx 0.5\,c$,
			which would strongly favor a sizeable transverse
			flow.
			}
		\end{minipage}
		\end{minipage}
	\end{minipage}
\end{figure}

We condense the information from the $m_T$-spectra with respect to the
transverse flow in Fig.~\ref{fit2all} where we show all
possible fit pairs $(T,\beta_s)$ for all hadronic spectra.
Without transverse flow ($\beta_s=0$) the intercepts of the curves with
the axis show the fitted temperatures of stationary thermal emitters
including resonance decays (Figs.~\ref{dmtpith} \& \ref{dmtallth}),
where all temperatures cluster around the value $T\approx 210\pm20\,\rm MeV$.
Surprisingly, despite their somewhat different shapes, the curves stay together
in one band and only weakly concentrate in the region $\beta_s\approx 0.5\,c$.

Intuitively one would rather expect the contrary:
Because flow increases the particle energies according to their rest masses
($\langle E \rangle \approx \langle E \rangle_{th} + {{m_0}\over{2}}
v^2_{coll}$),
the heavier particles profit more from the collective motion than the lighter
ones. Since both transverse flow and the local temperature determine the
slope of the $m_T$-spectrum, a bigger effect from flow for the heavier
particles should result in faster dropping $(T,\beta_s)$-curves from the fits
for $K^0_s$, $p$ and $\Lambda$. The different shapes of the curves
should make it possible to find one unique intersection point, representing
$T$ and $\beta_s$ of the collective system \cite{Search}.
Unfortunately, this simple argument does not take the resonance decays
into account, which modify especially the pion spectra strongly
\cite{Sollfrank}.
We show in Fig.~\ref{fit2allnoreso} the curves, which
unrealistically omit the resonance decays,
and from which we would erroneously deduce a transverse flow of
$\beta_s \approx 0.5\,c$ from the unique intersection of all curves.
However, as we see in Fig.~\ref{fit2all}, reality is more complicated,
the pion curve is considerably modified by resonance decays so that in the end
its shape resembles the heavy particles' and
all values $0<\beta_s<0.7\,c$ seem to be possible within the experimental
uncertainties.

The $\Lambda$ do not quite fit into the picture, they seem a bit colder
than the other particles including the protons.
An explanation in terms of an admixture of colder
$\Lambda$ from the fragmentation region is supported by the the rapidity
distribution but is in some contradiction to the higher proton temperature
(which however has experimentally the biggest error bars).
In a similar manner one could justify the higher temperatures
of the $\overline \Lambda$ which seem to originate more from the hotter center.

We can conclude that the data alone do not make a unique statement
about the existence of transverse flow. Instead of a single point for
$\beta_s$ there is a band with many pairs $(T,\beta_s)$.
The phenomenological analysis has fully exploited the available data
and has to stop here. Which point on these curves corresponds to the
freeze-out point of a consistent hydrodynamical evolution will be studied
in a seperate paper \cite{circumstantial,hydrodraft}.
The theoretical methods presented there can use the gathered information
to back-extrapolate towards the early stages of a heavy-ion collision and
to eventually fix the amount of transverse flow.

\section{Discussion And Conclusions}
\label{discussionandconclusions}

Naturally there is some doubt whether collective effects are actually
as important as it might seem from the analysis presented here, since the
$S$ nucleus is still very small. Comparing $S$+$S$ to $pp$ minimum bias data,
which should resemble most closely the average nucleon--nucleon collision
in $S$+$S$, we realize that the rapidity
distributions are rather similar in shape. Accounting for the isospin
symmetry, the width of the $S$+$S$-distribution turns out to be
only about 10\% smaller than from $pp$~\cite{NA35charged,Stroebele}.

The transverse momentum spectra show bigger differences when plotted
as the (normalized) ratio of $SS/pp$~\cite{WenigD}.
Apparently there is an enhancement
at low $p_T$, which could result from the increased population of resonances,
and a significant enhancement at high $p_T$, which we would contribute
to the bigger transverse flow. However, a high $p_T$-enhancement has also
been observed in $pA$ collisions \cite{Cronin} (Cronin effect),
although this effect sets in only above $p_\perp\approx 1.5\,\rm GeV$.
It has been interpreted as resulting from multiple collisions
of the partons in the nucleus~\cite{hardscatt} and as such
is a QCD-specific effect.
We interpret this multiple scattering already in $pA$ as
an indicator that collective matter flow can be generated, since in
the limit of many scatterings hydrodynamic behaviour will
eventually result.

It would be too impulsive to deduce from the apparent similarity of $pp$
and $S$+$S$ spectra that the collective interpretation is wrong, since
a) $pp$ is by no means an elementary collision system which
we understand in sufficient detail to serve as the antipode of a collective
system and
b) only a few of the observed features of the spectra can be
fully reproduced by these two radically different philosophies
which share only a small set of common principles as
local energy--momentum conservation, relativistic space--time picture, etc.
For other observables, which are not directly tied to the dynamics,
e.g. strangeness, we already see a big enhancement compared to $pp$, thus
indicating fundamental differences between both collision systems.

A careful systematic study of the $A$-dependence from $pp$ up to $Pb$+$Pb$
should allow to decide this question by showing the increasing importance
of the collective effects for the bigger systems relative
to the individual scattering processes.
Unfortunately at the present time the only other symmetric collisions
at high energies are $pp$ collisions \cite{NA5}, because, as we already
mentioned, we cannot directly compare with asymmetric ($pA$) collisions.
The data from the no longer operating ISR (Intersecting Storage Rings)
from light nuclei ($dd$, $\alpha\alpha$) \cite{ISR}
would also be very interesting,
but they probably have to be reanalysed for a direct comparison.

Nevertheless, from our experience with $S$+$S$ we can already risk
the prediction that the spectral shape from $Pb$+$Pb$ collisions
will show only gradual
changes from $S$+$S$ and that a careful analysis will be needed
to uncover the interesting physics. How much influence the creation
of a plasma would have on the spectra can not be answered in this context
without additional theory, but we would already like to caution big
expectations, since the plasma would be created in the early stages of the
collision and any signal would be leveled by the following hadronic stages
with their own dynamics.

In conclusion we have shown that a thermal model is perfectly possible
for $S$+$S$ collisions despite (or because of?) the similarity of
$S$+$S$ and $pp$ spectra. The data force us to include resonance decays
and longitudinal flow while they make no decisive statement about the
existence of transverse flow. But through further theoretical investigations
the quantitative analysis presented here can be used to prove the existence of
transverse flow and infer other informations about the collision zone
\cite{circumstantial,hydrodraft}.

\bigskip\noindent
{\sectionfont Acknowledgments}
\medskip

{\noindent \sloppy
We thank S.~Wenig for supplying us with the charged particle data
prior to publication.
E.~S.\ gratefully acknowledges support by the Deutsche Forschungsgemeinschaft
\newline
(DFG), the Alexander-von-Humboldt Stiftung and the U.S.\ Department of Energy
under contract number DE-AC02-76H00016.
J.~S.\ gratefully acknowledges support
by a fellowship from the Free State of Bavaria and the DFG.
U.~H.\ gratefully
acknowledges support by the DFG, the Bundesministerium f{\"u}r Forschung
und Technologie (BMFT), and the Gesellschaft f{\"u}r
Schwerionenforschung (GSI).
}

\begin{appendix}

\section{Slopes From Transverse Flow}
\label{slopestransverseflow}

Both transverse flow and resonance decays have a strong impact
on the $m_T$ spectra. In this and the following section we will show the
actual computation of the spectra, and we will try to
disentangle both influences by analysing the slope of the $m_T$-spectrum.

We follow the spirit of the boost invariant scenario by first defining
the transverse velocity field in the central slice
(using the boost angle $\rho = \tanh^{-1} \beta_r$) for azimuthal symmetry
\begin{equation}
{u'\,}^\nu( \tilde t, r, z=0 ) = (\cosh\rho, \,\vec e_r \sinh\rho, \,0 )
\end{equation}
and later boosting it in longitudinal direction (boost angle $\eta$) to
generate the whole velocity field \cite{Ruuskanen}
\begin{equation}
u^\mu(\rho,\eta)
= (\cosh\rho \,\cosh\eta, \,\vec e_r \sinh\rho, \,\cosh\rho\,\sinh\eta ) \;.
\end{equation}
Note that $u^\mu$ is not symmetric with respect to the
boost angles $\rho$ and $\eta$. This results from the fact that Lorentz
transformations for different directions do not
commute.

Having defined the velocity field, the invariant momentum spectrum is
given by \cite{CooperFrye}:
\begin{equation}
\label{Ed3ndp3CooperFrye}
E{{d^3n}\over{d^3p}}
= \int_\sigma f(x,p)\, p^\lambda d\sigma_\lambda
\approx {{g}\over{(2\pi)^3}} \int e^{-(u^\nu p_\nu -\mu )/T}
				p^\lambda d\sigma_\lambda	\;,
\end{equation}
where $f(x,p)$ is the invariant distribution function, which we assume
to be an isotropic thermal distribution boosted
by the local fluid velocity $u^\mu$, and we approximate the respective
Bose and Fermi distributions by the Boltzmann distribution.

This mathematical formulation measures directly the particle flow through
the given hypersurface $\sigma$ as if the virtual walls of the fluid cells
have suddenly disappeared and the particles are flying
isotropically in all directions. $\sigma$ should not be visualized as a real
surface which emits radiation only to the outside. Instead it defines
the borderline between hydrodynamical behaviour and free-streaming particles,
which are both idealizations, as a mathematical construct.
More intuitive approaches lead to formulas~\cite{Sinyukov} which no longer
obey the conservation laws. In reality the freeze-out hypersurface
might be defined by the points of the last interaction of each
individual particle and thus acquires a thickness of the order
of the mean free path. This seems to be difficult to implement
consistently with the dynamical evolution and
for our purposes the simple model seems to be quite sufficient.

We parametrize the hypersurface $\sigma(r,\phi,\zeta)$ in cylindrical
coordinates $0\le r\le R$, $0\le \phi <2\pi$ and $-\Zeta\le\zeta\le\Zeta$
in longitudinal directions. We invoke instantaneous freeze-out
in $r$-direction to keep matters simple, but allow for more general shapes
$\big(t(\zeta),z(\zeta)\big)$
in longitudinal direction (e.g.\ hyperbolas \cite{Bjorken}) because of the
large longitudinal time dilatation effects:
\begin{eqnarray}
\sigma^\mu(r,\phi,\zeta)
&&= \big(\, t(\zeta),\, r\cos\phi,\, r\sin\phi,\, z(\zeta)\, \big)
\\
p^\mu d\sigma_\mu
&&= \left[  E   {{\partial z}\over{\partial \zeta}}
	  - p_L {{\partial t}\over{\partial \zeta}} \right]
   rdr\,d\phi\,d\zeta \quad.
\end{eqnarray}
For the argument of the Boltzmann distribution we need the momentum of the
particle in the center-of-fireball system
\begin{eqnarray}
u^\mu p_\mu
&=& m_T \cosh( y-\eta) \cosh\rho - p_T \sinh\rho \cos(\phi-\varphi) \;.
\end{eqnarray}

Because of azimuthal symmetry we can integrate over $\phi$ making use of
the modified Bessel function
$I_0(z) = (2\pi)^{-1} \int_0^{2\pi} e^{z\cos\phi} d\phi$:
\begin{eqnarray}
E{{d^3n}\over{d^3p}}
&=& {g\over{(2\pi)^2}} \int_{-\Zeta}^\Zeta d\zeta
	\left[ m_T \cosh y {{\partial z}\over{\partial \zeta}}
	- m_T \sinh y {{\partial t}\over{\partial \zeta}} \right] \\
&\times& \int_0^R rdr\,
		\exp\left(-{{m_T \cosh\rho\cosh(y-\eta)-\mu}\over{T}}\right)
		I_0\left( {{p_T \sinh\rho}\over{T}}\right)
\nonumber
\end{eqnarray}

For the transverse mass spectrum we integrate with the help of another modified
Bessel function
$K_1( z )=\int_0^\infty \cosh y \, e^{-z\cosh y} dy$:
\begin{eqnarray}
\label{dndmttransflow}
{{dn}\over{m_T dm_T}}
\!\!\!\!&=&\!\!\!\! {g\over\pi} m_T
	\!\!\!\int_{-\Zeta}^\Zeta \!\!\!\!d\zeta
		\left [ \cosh\eta {{\partial z}\over{\partial \zeta}}
		       -\sinh\eta {{\partial t}\over{\partial \zeta}} \right]
	\!\int_0^R \!\!\!\!rdr\, K_1\left( {{m_T \cosh\rho}\over{T}}\right)
				    I_0\left( {{p_T \sinh\rho}\over{T}}\right)
\nonumber \\\!\!\!\!&=&\!\!\!\! {2g\over\pi} m_T
	Z_{\tilde t}
	\int_0^R \!\!\!\!rdr\, K_1\left( {{m_T \cosh\rho}\over{T}}\right)
				    I_0\left( {{p_T \sinh\rho}\over{T}}\right)
\end{eqnarray}

We see that the transverse mass spectrum factorizes, as long as temperature
and transverse flow are independent of the longitudinal position in a
longitudinally comoving coordinate system. There is only a factor
$Z_{\tilde t}$ affecting the normalization.
The invariant momentum spectrum factorizes only for
small transverse flows ($\cosh\rho\approx 1$)
into a longitudinal and a transverse part. The rapidity distribution
is almost independent of transverse flows, even if the latter becomes
large, as we have checked explicitly.

The general formula for azimuthal symmetry
allowing non-instantaneous freeze-out in $r$-direction
($\partial t/\partial r \not = 0$) is rather similar,
involving also $K_0$ and $I_1$:
\begin{eqnarray}
\label{dndmttransflowgeneral}
{{dn}\over{m_T dm_T}}
\!\!\!\!&=&\!\!\!\! {g\over\pi}
	\!\int_{-\Zeta}^\Zeta \!\!\!\!d\zeta
	\!\int_0^{R(\zeta)} \!\!\!\!rdr \left\{		\left [ \cosh\eta {{\partial
z}\over{\partial \zeta}}
	           -\sinh\eta {{\partial(t,r)}\over{\partial(\zeta,r)}} \right]
	\right.\nonumber\\
	&&\qquad\qquad\qquad\qquad\qquad	m_T K_1\left( {{m_T
\cosh\rho}\over{T}}\right)
				    I_0\left( {{p_T \sinh\rho}\over{T}}\right)
\nonumber\\&&\qquad\qquad\qquad	\left.	- {{\partial z}\over{\partial
\zeta}}{{\partial t}\over{\partial r}}	p_T K_0\left( {{m_T
\cosh\rho}\over{T}}\right)
			    I_1\left( {{p_T \sinh\rho}\over{T}}\right)
\right\}\nonumber\\\end{eqnarray}

We can achieve a better understanding of the transverse mass spectrum
in (\ref{dndmttransflow}) by analysing its slope. Fixing the variables
$r$, $\eta$, $T$ at reasonable values we focus on the $m_T$ dependent parts.
The slope in a semilogarithmic plot is
\begin{eqnarray}
  {{d}\over{dm_T}} \ln \Bigl( {{dn}\over{m_Tdm_T}} \Bigr)
&=&{{d}\over{dm_T}} \ln \Bigl( m_T I_0\Bigl({{p_T\sinh\rho}\over{T}}\Bigr)
				    K_1\Bigl({{m_T\cosh\rho}\over{T}}\Bigr)
				    \Bigr)
\nonumber\\
&=&
{{I_1\bigl({{p_T\sinh\rho}\over{T}}\bigr)}
	\over{I_0\bigl({{p_T\sinh\rho}\over{T}}\bigr)}}
		{{m_T}\over{p_T}} {{\sinh\rho}\over{T}}
  -
{{K_0\bigl({{m_T\cosh\rho}\over{T}}\bigr)}
\over{K_1\bigl({{m_T\cosh\rho}\over{T}}\bigr)}}
		{{\cosh\rho}\over{T}}
\end{eqnarray}

\begin{figure}[htb]
	\begin{minipage}[b]{\medfigsize}
		\epsfxsize \medfigsize \epsfbox{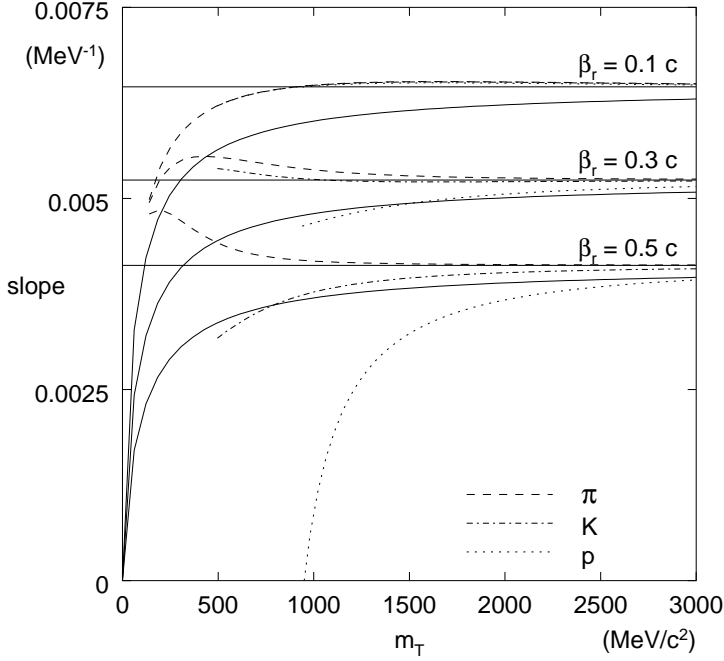}
	\end{minipage}
	\hfill
	\begin{minipage}[b]{\medcapsize}
		\caption[]{\label{slopedndmt} \sloppy
			{\bf \boldmath \sloppy $m_T$-slopes
			with transverse flow \unboldmath}
			for pions, kaons and protons.
			Every group of curves corresponds to a fixed
			transverse expansion velocity $\beta_r$.
			The approximations of the curves
			by the pure exponentials (horizontal lines)
			are remarkably good already at low $m_T$.
			For comparison we included also the slopes of thermal
			sources (in Boltzmann statistics)
			with correspondingly
			blue shifted temperatures (solid curves).
			Compare also with Fig.~\ref{resoslope}.
			}
		\vspace{0.5 truecm}
	\end{minipage}
\end{figure}

We can immediately derive the limiting case for large $m_T$, since there
$m_T/p_T \to 1$,  $K_0 / K_1 \to 1$ and for finite flow
($\sinh\rho>0$) also $I_1 / I_0 \to 1$:
\begin{equation}
\label{blueshift}
\lim_{m_T\to\infty} {{d}\over{dm_T}} \ln \Bigl( {{dn}\over{m_Tdm_T}} \Bigr)
= {{\cosh\rho-\sinh\rho}\over{T}}
= - {1\over T} \sqrt{{{1-\beta_r}\over{1+\beta_r}}}
\end{equation}

The apparent temperature, understood as the inverse slope at high $m_T$,
is then larger than the original temperature by a blue shift factor, see
also~\cite{Gersdorffhydro}:
\begin{equation}
T_{\rm eff} = T \sqrt{ {1+\beta_r}\over{1-\beta_r} }
\end{equation}

For small values of $m_T$ the situation is much less clear, as one can see
from Fig.~\ref{slopedndmt}. We cannot expect a simple expansion around
$m_T=\infty$ to deliver fully satisfying results, because apart from the
shape of the Bessel functions also the particle mass $m_0$
(via $p_T = \sqrt{{m_T}^2 - {m_0}^2}$) and the fluid rapidity $\rho$
influences the slope. Nevertheless, the basic tendency can be discussed
with this expansion
\begin{equation}
{{d}\over{dm_T}} \ln \Bigl( {{dn}\over{m_Tdm_T}} \Bigr)
\approx - {{\sinh\rho-\cosh\rho}\over{T}}
	- {{m_0^2}\over{2m_T^2}} {{\sinh\rho}\over T}
	+ {{3}\over{16}} {{T}\over{m_T^2}}
		\Bigl( {{1}\over{\sinh\rho}} + {{1}\over{\cosh\rho}} \Bigr)
	\qquad
\end{equation}
The first term is again the blue shift factor, the additional
terms are responsible for the curvature of the spectrum.
The deviations from the
exponential behavior are largest for large flows ($\sinh\rho\gg 0$)
and have the biggest effect at low $m_T$, where the shape of the Bessel
functions tends to steepen the spectra, while big particle masses strongly
flatten the spectrum there \cite{Search}.

One can extract regions from the figure, where the slope approaches the
limiting value satisfactorily. For pions and kaons we have
$m_T> 1000\,\rm MeV$, whereas for nucleons in the case of strong flow
the limiting slope will be approached
only above $m_T>2000\,\rm MeV$. This has to be contrasted with the behavior
of the resonance decays which only affect the low $m_T$ region
(see Fig.~\ref{resoslope}).

Since in a realistic picture there are various components of transverse flow,
we account for them by a superposition of simple exponentials,
each with its own blue shift, denoted by
$b_r(\beta_r) = \sqrt{ (1-\beta_r) / ( 1+\beta_r) }$, to compute the
blue shift factor of the integrated spectrum.
\begin{equation}
b_{\rm int} = -T \cdot {{d}\over{dm_T}} \ln \int_0^R
	\exp\Bigl( - b_r(\beta_r) {{m_T}\over{T}} \Bigr)
	r dr
\end{equation}

The integrals could be solved analytically for $n=1$ and $2$, but it is more
instructive to expand them into a power series in $1-b_s$ under the assumption
that $m_T/T$ is not too big:
\begin{equation}
\label{eqslopeapp}
b_{\rm int} \approx \left\{
\begin{array}{*{4}{r@{\,}}r@{\;}l}
1\!     &-{1\over{2}} (1 \!-\! b_s)
	&-{{ (m_T/T)-3}\over{12}} (1 \!-\! b_s)^2
	&+\ldots
	&&(n=2) \\
1\!     &-{2\over 3} (1 \!-\! b_s)
	&-{{2(m_T/T)-5}\over{36}} (1 \!-\! b_s)^2
	&+\ldots
	&&(n=1) \\
1\!     &-{4\over 5} (1 \!-\! b_s)
	&-{{2(m_T/T)-3}\over{75}} (1 \!-\! b_s)^2
	&+\ldots
	&&(n={\textstyle{1\over2}}) \\
1\!     &-(1 \!-\! b_s)
	&
	&
	&&(n=0)
\end{array}
\right.
\end{equation}
where the terms in third order do not improve the approximation considerably.
Approximating the blue shift factor to first order by $b_s \approx 1-\beta_s$,
we arrive at a surprising coincidence with the average fluid velocity
\begin{equation}
b_{\rm int}
  \approx \sqrt{ {{1-\langle\beta_r\rangle}\over{1+\langle\beta_r\rangle}} }
  \approx 1- {2\over{n+2}} \beta_s
\end{equation}
which is a handy way to estimate the behaviour of the fits in
Fig.~\ref{fit2all}.

In conclusion we find from our simple quantitative estimates
that the influence of the transverse flow on the
transverse momentum spectrum is strong. We also investigated the influence of
the geometry of the freeze-out
hypersurface on both the $y$- and on the $m_T$-spectra.
General formulas developed in \cite{Diplom,Diss} show that
the shape of the hypersurfaces influences the spectra via the partial
derivatives of $\sigma$ with respect to $r$ and $z$.
Similarly to the spherical case~\cite{Search}
we have not encountered big differences in the width of the $y$-spectra
as well as the slope of the $m_T$ spectra
as long as we confined ourselves to realistic freeze-out hypersurfaces.

\section{Slopes From Resonance Decays}
\label{slopesresonances}

Computing the spectrum of the resonance products is generally
a complex task and requires a computer to do the phase space integrals
numerically. However, investigating the simpler case of the two-body decay
analytically can give us some insights into the resulting spectra. We start
from the general formula in \cite{Sollfrank}, now specialized for
2-body decays:
\begin{eqnarray}
{{d^2n}\over{dy m_T dm_T}}
&=& {{m_R b}\over{4\pi p^*}}
	\int\limits_{y_R^{(-)}}^{y_R^{(+)}}
	{{dy_R}\over{\sqrt{ m_T^2 \cosh^2(y-y_R) - p_T^2} }}
\\ &&\qquad\times
	\int\limits_{m_{T R}^{2\,(-)}}^{m_{T R}^{2\,(+)}}
	{{dm_{T R}^2}\over{\sqrt{ (m_{T R}^{(+)} - m_{T R})
			 (m_{T R} - m_{T R}^{(-)}) } }}
	{{d^2n_R}\over{dy_R m_{T R} dm_{T R} }}
\nonumber
\end{eqnarray}

\begin{figure}[tb]
	\begin{minipage}[b]{\medfigsize}
		\epsfxsize \medfigsize \epsfbox{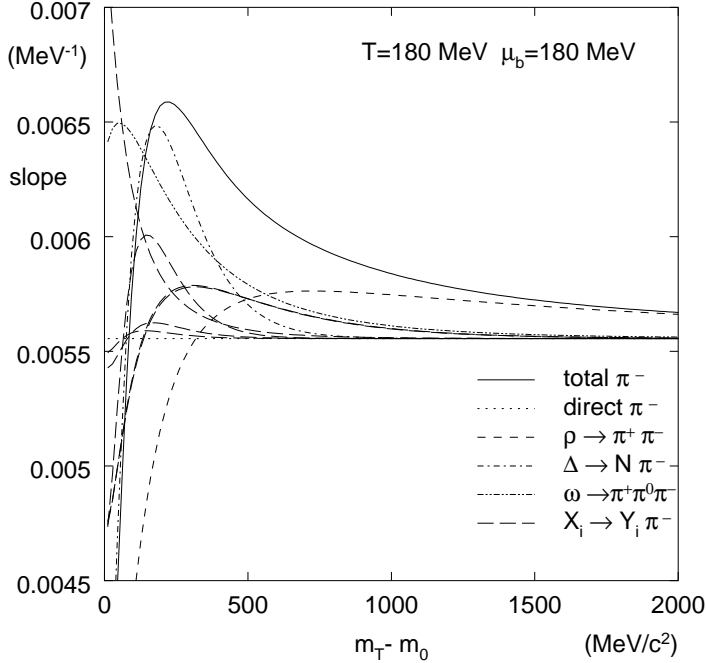}
	\end{minipage}
	\hfill
	\begin{minipage}[b]{\medcapsize}
		\caption[]{\label{resoslope} \sloppy
			{\bf \boldmath $m_T$-slopes of the $\pi^-$
			with resonance decays.	\unboldmath}
			For realistic values of $T$ and $\mu_b$ we show
			the slopes resulting from the addition of always one
			resonance channel to the purely thermal $\pi^-$-spectrum.
			At high $m_T$ only the $\rho$ decay changes the
			slope by a few percent. At lower $m_T$ many individual
			contributions (especially $\Delta$ and $\omega$) add up
			to a significant modification of the total spectrum.
			Interesting is the comparison with
			Fig.~\ref{slopedndmt} (scales are different).
			}
		\vspace{0.5 truecm}
	\end{minipage}
\end{figure}

By introducing a thermal spectrum for the resonances which is independent
of rapidity
${{d^2n}\over{dy \,m_T dm_T}} = C_1 m_{T R} e^{-m_{T R}/T}$,
we can solve one further integral analytically:
\begin{eqnarray}
{{d^3n}\over{dy m_T dm_T}}
&\!\!=\!\!& {{m_R}\over{4\pi p^*}}
	\int\limits_{y_R^{(-)}}^{y_R^{(+)}}
		{{dy_R}\over{\sqrt{m_T^2\cosh^2(y-y_R)-p_T^2} }}
\\ &&\qquad\times
	\pi C_1 e^{-a/T} \left\{ 2(a^2+b^2) I_0\Big({b\over T}\Big)
				- (4ab+2Tb) I_1\Big({b\over T}\Big)
			\right\}
\nonumber
\end{eqnarray}
where $a ={1\over 2} (m_{T R}^{(+)} + m_{T R}^{(-)})$ and
$b={1\over 2} (m_{T R}^{(+)} - m_{T R}^{(-)})$.
For large $m_T$ $a$ and $b$ are approximately independent of $y_R$,
\begin{equation}
a\approx {{m_R E^* m_T}\over{m^2}} := \bar a \quad,\qquad
b\approx {{m_R p^* p_T}\over{m^2}} := \bar b
\end{equation}
and we can extract in first order the slope of the transverse mass
spectrum since the Bessel functions $I_i(\bar b/T)$ behave asymptotically
like exponentials $\exp(\bar b/T)$.
The resulting exponential is $\exp((\bar b-\bar a)/T)$
giving a slope at high $m_T$
\begin{equation}
\label{resoapprox}
{{d}\over{dm_T}} \ln {{dn}\over{m_T dm_T}} \approx
	- {{m_R (E^*-p^*)}\over{m^2}} {1\over T} \qquad.
\end{equation}

$E^*$ and $p^*$ are energy and momentum of the daughter particle in the
rest system of the resonance and are for the two body decay already given by
the
particle masses ($p^*$ is listed for every decay channel
in~\cite{particledatagroup}). The estimate (\ref{resoapprox})
coincides very well with the slopes of the numerically computed
spectra at large $m_T$.

Further approximations reveal for the slope
${{m_R}\over{p^*}} {1\over T}$ or th effective temperature
\begin{equation}
T_{\rm eff} \approx {{p^*}\over {m_R}} T
\end{equation}
Since $p^* < m_R$ the spectra
of the daughter particles are generally steeper than the original
spectrum of the resonance, which translates into a lower apparent temperature
of the pion spectra at low $m_T$ (Fig.~\ref{resoslope}).

We can also estimate the width of the rapidity distribution by looking
at the integration limits for $y_R$. The maximal rapidity difference
between resonance and daughter particle is restricted by kinematics to
\begin{equation}
\vert y_R - y \vert \le \sinh^{-1} \left({{p^*}\over{m_T}} \right)
\end{equation}
For the $\Delta$-decay $p^* = 227\rm\,MeV$, so that the decay pion can move
at most 1.27 units of rapidity (at $m_T=m_0$), and the nucleon at most 0.24.
The actual $(y_R-y)$-distribution of the daughter particle will naturally
be much smaller, as can be inferred from the width of an
isotropic distribution, which is for massless particles
$\Gamma^{\rm fwhm} = 2 \times 0.88$ (eq.~\ref{dndythermal}). This similarity to
a thermal source
also explains the small influence of the resonance decays on the {\it shape}
of the rapidity distribution (Fig.~\ref{dypi}), which is only scaled by the
decay contributions.

In conclusion we have shown why the decay spectra are concentrated in the
low $m_T$ region. However, because of
the superposition of many channels and the complex structure of the
decay spectrum itself, we cannot estimate the influence
by an easy method and have to resort to the full numerical
evaluation of (\ref{dymtdmtreso}).

\end{appendix}


\end{document}